\documentclass[11pt]{article}

\usepackage[margin=1.1in]{geometry}
\usepackage[T1]{fontenc}
\usepackage[utf8]{inputenc}
\usepackage{lmodern}
\usepackage{microtype}
\usepackage{booktabs}
\usepackage{amsmath}
\usepackage{xcolor}
\usepackage{listings}
\usepackage{multirow}
\usepackage[hidelinks]{hyperref}
\usepackage{caption}
\lstdefinestyle{code}{
  basicstyle=\ttfamily\small,
  breaklines=true,
  frame=single,
  framerule=0.3pt,
  xleftmargin=1em,
  aboveskip=0.8em,
  belowskip=0.8em,
  columns=fullflexible,
  keepspaces=true
}

\newcommand{\jssam}{JS-SAM}
\newcommand{\tlaplus}{TLA\texorpdfstring{\textsuperscript{+}}{+}}
\newcommand{\barenext}{bare-\texttt{next}}

\title{\textbf{Executable JavaScript as a Checkable Specification Language:\\
A \jssam{} Case Study on SysMoBench}}

\author{
  Jean-Jacques Dubray\\
  \texttt{jdubray@gmail.com}\\[0.5em]
  \small\textit{In collaboration with the SysMoBench / Specula team}
}
\date{July 13, 2026}

\begin{document}
\maketitle

\begin{abstract}
Can large language models write faithful formal specifications of real
systems, and does it matter whether they write in a formal language they
have seen rarely or in a mainstream language abundant in their training
data? We study this on SysMoBench, which grades a generated specification in
four phases, the decisive one replaying execution traces captured from the
running system. We add \jssam{}, its first non-formal backend, in which a
specification is executable JavaScript written in the SAM pattern, a pattern
whose semantics mirror \tlaplus{}, and run a controlled comparison that
separates three variables an ordinary head-to-head entangles: the language,
the specification contract (the shape the model must fill), and the prompt.
The study spans four frontier models and three systems, an operating-system
spinlock, a distributed lock service, and the Etcd Raft consensus
implementation, with counterexample-driven repair.

Three findings emerge. First, conformance against the real system is the
only phase that discriminates among models; internal consistency is
inexpensive to satisfy, and a specification that looks right is not thereby
right. Second, once the comparison is drawn like for like, the specification
contract, not the language, governs fidelity: JavaScript in the shape of the
\tlaplus{} transition relation is as faithful as \tlaplus{}. Third, a minimal
contract carries transcription but not semantic derivation: at consensus
scale the difficulty becomes understanding the protocol, which no contract
shape and no language supplies. We frame executable JavaScript as a
\emph{checkable} specification substrate that complements, rather than
replaces, the verification \tlaplus{} provides, and present the study as a
case study.
\end{abstract}

\section{Introduction}
\label{sec:intro}

Formal specifications are among the most effective tools for establishing the
correctness of concurrent and distributed systems~\cite{newcombe2015amazon},
and among the least used: writing them demands rare expertise and sustained
effort. LLMs raise an obvious question, namely whether they can write
specifications for us, and a harder one: \emph{how would we know if they got
it right?} A generated specification can be syntactically flawless, explore
cleanly under a model checker, satisfy plausible invariants, and still
describe a system that does not exist. The SysMoBench benchmark~\cite{sysmobench}
confronts this directly. Its authors observed that when asked to model Etcd's
Raft implementation, a frontier LLM produced, in effect, the specification
from the appendix of the Raft paper, a ``textbook'' model with little
connection to the implementation at hand~\cite{sigops2026}. SysMoBench
therefore grades specifications in four phases of increasing demand:
(1)~syntax, (2)~runtime model checking, (3)~\emph{transition validation},
which replays $(\mathit{pre}, \mathit{action}, \mathit{post})$ windows
captured from real executions against the generated model, and
(4)~invariant verification.

SysMoBench was built for \tlaplus{}~\cite{lamport2002specifying}, and has since
grown a pluggable backend interface with Alloy~\cite{jackson2002alloy} and
PAT/CSP\#~\cite{sun2009pat} backends. All three are formal languages, and all
three are rare in LLM training corpora relative to mainstream programming
languages. This asymmetry motivates the extension we present here.

\paragraph{The \jssam{} hypothesis.}
SAM~\cite{sam,dubray2016infoq} is a software-engineering pattern whose
semantics are explicitly derived from \tlaplus{}: \emph{actions} compute
proposals from inputs, a \emph{model} accepts or rejects proposals as the sole
locus of mutation, and \emph{state} is a pure function of the model. A SAM
specification is an ordinary executable JavaScript module. JavaScript is
plausibly the single most abundant programming language in LLM training data.
The hypothesis, then:

\begin{quote}
\emph{Can an LLM produce a correct specification of a real system more readily
in a pattern hosted by a language it has seen constantly in training
(JavaScript), than in formal languages it has rarely seen, while preserving
the \tlaplus{}-style discipline that makes the artifact checkable?}
\end{quote}

Either answer is informative. If \jssam{} scores diverge upward, host-language
familiarity dominates and formal syntax is a bottleneck; if scores match or
fall, the hard part is modeling, not notation.

\paragraph{Contributions.}
This paper reports the first end-to-end results from the \jssam{} backend,
all on the SysMoBench \texttt{spin} task (the Asterinas OS
spinlock~\cite{asterinas}):

\begin{enumerate}
\item \textbf{A non-formal-language backend for SysMoBench}
  (\S\ref{sec:backend}): the \jssam{} module contract, a sandboxed execution
  helper, and, a first for the benchmark, a \emph{direct} Phase-3
  replay path requiring no coding-agent orchestration, because SAM's step
  relation \emph{is} $\mathit{model.present}(\mathit{action}(\mathit{data}))$.
\item \textbf{A reproducible kernel trace-capture methodology}
  (\S\ref{sec:traces}): instrumenting the real Asterinas spinlock, driving a
  two-thread contention scenario in a single-CPU kernel test without deadlock,
  and folding low-level kernel events into model-independent
  $(\mathit{pre}, \mathit{action}, \mathit{post})$ windows, producing the
  28-window corpus used throughout.
\item \textbf{A four-model comparison} (\S\ref{sec:models}) showing that
  Phases 1, 2, and 4 have no discriminating power on this task while Phase 3
  spreads the models from 0\% to 86.4\% on state-changing windows (21.4\% to
  89.3\% headline, against an identity base rate of 21.4\%), and that
  general model capability does not predict modeling accuracy: the most
  capable model tested produced a specification that no-ops every lock
  release.
\item \textbf{A repair-loop experiment} (\S\ref{sec:repair}) demonstrating
  that self-correction from precise trace feedback is a distinct capability
  axis from one-shot accuracy: capable models repair fully (verified by a
  held-out generalization audit as fixes, not memorization), the weakest not
  at all, with mid-tier outcomes unstable at $N{=}1$.
\item \textbf{A \jssam{}-vs-\tlaplus{} evaluability comparison}
  (\S\ref{sec:vstla}): 4/4 \jssam{} specifications reach end-to-end
  evaluability versus 1/4 for \tlaplus{}, reported strictly per cause
  (a known TLC idiom trap twice, a harness fault once) and shown to vanish
  under a one-line prompt guardrail: a default-prompt robustness
  observation, not a language-capability finding.
\item \textbf{A controlled cross-language conformance study}
  (\S\ref{sec:crosslang}): a pre-registered, paired design realized as a
  full $3{\times}2$ factorial in (contract, prompt)
  ($N{=}5$ generations per cell), pinning both languages
  to the same observable-state contract and replaying the same trace corpus
  mechanically in both. The completed factorial attributes the conformance
  gap to the SAM contract's machinery, not to the JavaScript language, not
  to formality, and not (as an incomplete design had suggested) primarily to
  prompt prescriptiveness: the \barenext{} contract reaches 100\% for every
  model \emph{without} the semantics block, while the semantics block within
  SAM moves only two of four models. This yields a concrete design change for
  the backend, validated by 40/40 \barenext{} specifications passing every
  phase across two tasks (\S\ref{sec:locksvc}).
\item \textbf{A consensus-class boundary study} (\S\ref{sec:raft}): the
  \barenext{} contract carried to Etcd Raft, a 97-window corpus captured
  through the implementation's own trace instrumentation, validated by a
  hand-written reference specification at 100\%, where 0 of 20
  one-shot \barenext{} specifications pass every phase. The failure analysis
  separates what the contract still buys (transcription, evaluability)
  from what it cannot (semantic derivation, and a specification-vs-program
  distinction the strongest model reasons its way past), and a one-round
  repair experiment stratifies the four models into convergent, slowly
  convergent, regressing, and plateaued.
\item \textbf{An ecological check on the deployed \tlaplus{} path}
  (\S\ref{sec:agentpath}): the benchmark's agent-mediated Phase 3, run on
  the Experiment-3 specifications, confirming the harness-vs-model
  attribution of the evaluability gap and exposing a scoring-semantics
  difference between the deployed and controlled oracles.
\end{enumerate}

In light of the results, we regard the trace-capture methodology (2), the
controlled-comparison methodology (6), and the contract-shape finding as the
primary contributions; the \jssam{} backend (1) is the vehicle that surfaced
them, and its own results argue for the \barenext{} revision of its contract.

\paragraph{Scope of claims.}
Four boundaries govern everything that follows, and we state them before
the results rather than after.
(i)~\emph{Checkable, not verifiable.} What this paper demonstrates is that
executable JavaScript specifications can be \emph{checked}: bounded
safety exploration over reachable states, and conformance replay against
finite execution traces. It does not demonstrate verification in the sense
\tlaplus{} enables: no liveness or temporal-logic properties, no fairness,
no unbounded state spaces, no mechanical proof
(\S\ref{sec:scope}).
(ii)~\emph{Task scale.} The controlled factorial concerns one system,
the spinlock, a 28-window corpus, with the \barenext{}-contract validation
extending to a second task (a distributed lock service, 60 windows) and
to a third at consensus scale (Etcd Raft, 97 windows;
\S\ref{sec:raft}). The Raft study tests only the \barenext{} contract: the
factorial's SAM and \tlaplus{} arms remain spinlock-only, so the
contract-vs-prompt ordering is still conditional on the small tasks.
(iii)~\emph{Controlled and ecological, reported separately.} The
cross-language comparison uses a direct, mechanical TLC replay; the
benchmark's deployed agent-mediated \tlaplus{} Phase 3 was run
on the same specifications (\S\ref{sec:agentpath}) and confirms the
per-cause attribution, but the two paths score different window sets under
different oracles and their numbers are not interchangeable
(\S\ref{sec:scope}). The comparison is to TLC specifically, not the wider
\tlaplus{} toolchain.
(iv)~\emph{Limited statistical resolution.} Experiments 1--3 are
single-generation; the factorial runs $N{=}5$ per cell, where the exact
permutation test's smallest attainable $p$ is $2/252\approx0.0079$:
effect sizes, not $p$-values, are the primary quantities throughout.
We present the study as a case study whose value lies in the precision of
the failure attribution and the reproducibility of the pipeline, not as a
settled verdict on the hypothesis (\S\ref{sec:limitations}).

\section{Background}
\label{sec:background}

\subsection{SysMoBench}

SysMoBench~\cite{sysmobench} pairs eleven real systems (concurrent
synchronization primitives and distributed protocols, from the Asterinas
spinlock to Etcd Raft) with instrumentation harnesses and expert-written
invariants. Given the system's source, an LLM generates a specification, which
is scored in four automated phases:

\begin{description}
\item[Phase 1, Syntax.] Does the specification parse and structurally
  validate? (For \tlaplus{}: the SANY analyzer.)
\item[Phase 2, Runtime model checking.] Does bounded state-space
  exploration terminate without errors, deadlocks, or nondeterminism? (For
  \tlaplus{}: TLC~\cite{yu1999tlc}.)
\item[Phase 3, Transition validation (conformance).] Execution traces from
  the \emph{real} system are cut into windows
  $(\mathit{pre}, \mathit{action}, \mathit{post})$; each window passes if the
  generated model, started at $\mathit{pre}$ and applying $\mathit{action}$,
  produces $\mathit{post}$. The output is a per-action scorecard.
\item[Phase 4, Invariant verification.] Expert invariants (e.g., mutual
  exclusion) are translated into the specification's language and checked.
\end{description}

The benchmark's central published finding is that frontier LLMs cluster near
100\% on syntax but average roughly 46\% on conformance and 41\% on
invariants~\cite{sigops2026}: models write valid \tlaplus{} fluently yet
describe the textbook protocol rather than the implementation. Phase 3 exists
precisely to distinguish ``writing \tlaplus{}'' from ``modeling this system.''

\subsection{The SAM pattern}

SAM (State-Action-Model)~\cite{sam,dubray2016infoq} factors a program's
temporal behavior the way \tlaplus{} factors a specification. An
\emph{action} is a pure function from inputs to a \emph{proposal}; the
\emph{model} alone decides whether to accept a proposal and mutate itself
(the analogue of a \tlaplus{} next-state relation, with acceptors playing the
role of guards); \emph{state} is a pure function of the model. A step of the
system is $\mathit{model.present}(\mathit{action}(\mathit{data}))$. The
pattern is implemented by the \texttt{sam-pattern}
library~\cite{sampattern}, which also provides a bounded explorer
(\texttt{checker}) over the model's reachable states; in the \jssam{}
backend, this checker is what executes Phases 2 and 4.

Two properties make SAM suitable as a specification substrate. First, the
semantic alignment with \tlaplus{} means the benchmark's phases map onto it
without distortion: bounded exploration is Phase 2, replaying a trace window
is literally one SAM step, and invariants are predicates over
$\mathit{getState}()$. Second, the artifact is an executable JavaScript
module: there is no separate tool chain between the specification and its
own evaluation.

\section{The \jssam{} backend}
\label{sec:backend}

SysMoBench backends implement a \texttt{LanguageBackend} interface and
register themselves; the CLI and the four evaluators resolve them by name with
no further plumbing. The \jssam{} backend consists of a Python adapter and a
small Node.js helper (\texttt{tools/js-sam/cli.mjs}) with four subcommands
(\texttt{validate}, \texttt{check}, \texttt{transitions},
\texttt{invariants}), one per phase.

\paragraph{Module contract.}
The generation prompt instructs the model to emit a single CommonJS module:

\begin{lstlisting}[style=code]
module.exports = {
  instance,       // SAM instance (synchronous steps)
  init,           // () => void : reset to the initial state
  actions,        // { AcquireLock: (data)=>void, ReleaseLock: (data)=>void }
  getState,       // () => plain JSON-serializable snapshot
  setState,       // (snapshot) => void : force a state (Phase 3 replay)
  checkerIntents, // action descriptors + input domains (Phases 2 & 4)
};
\end{lstlisting}

\noindent
\texttt{actions} must cover exactly the task's target actions;
\texttt{getState} must return a plain object (this is what trace states are
diffed against); \texttt{setState} exists solely so Phase 3 can replay a
window without searching for a path to its pre-state; and the module must be
deterministic and I/O-free. \texttt{checkerIntents} declares each action's
finite input domain (for \texttt{spin}, every
$\{\mathit{thread}, \mathit{callType}\}$ combination), playing the role of a
\tlaplus{} constants block. Notably, there is \emph{no separate configuration
artifact}: the module is self-contained, a fact that becomes material in
\S\ref{sec:vstla}.

\paragraph{Sandboxing.}
Model-generated JavaScript is untrusted. Every helper invocation runs in a
throwaway \texttt{node:20-slim} Docker container with no network,
a read-only root filesystem, a non-root user, and CPU/memory/PID caps. The
container is the trust boundary.

\paragraph{The direct Phase-3 path.}
For \tlaplus{}, SysMoBench's transition validation drives TLC through a
coding-agent workflow that constructs a checking module per window. \jssam{}
needs none of that: the evaluator loads the module, and for each window
executes \texttt{setState(pre)}; \texttt{actions[name](data)}; and diffs
\texttt{getState()} against \emph{post} under the benchmark's projection rule
(only keys present in the trace's post-state must match, so model-internal
bookkeeping is ignored). This ``direct path'' is the first non-agent Phase-3
implementation in the benchmark, and it follows from SAM's semantics: a
window \emph{is} a SAM step.

\paragraph{Cross-platform engineering.}
Making the pipeline run on a Windows host (in addition to Linux/macOS)
required fixing several host assumptions in the harness (POSIX-only
\texttt{os.getuid()} calls, host-path container mounts invalid for
\texttt{C:\textbackslash} paths, and console encoding), plus threading the
helper's \texttt{stepsExplored} metric through a field previously hardcoded to
zero for non-\tlaplus{} backends. We also generalized the benchmark's
agent-based invariant translator (previously \tlaplus{}-only) into a shared,
language-neutral component that \jssam{} can use, although all experiments
below use the direct translator for parity across languages. These changes
are upstreamed as pull requests.

\section{Capturing ground truth: kernel traces for Phase 3}
\label{sec:traces}

Phase 3 is the only phase requiring data external to the model, and SysMoBench
does not ship trace corpora: traces are generated on demand by instrumenting
the real system. Additionally, the maintainers' own \texttt{spin} captures use
three threads while the \jssam{} prompt models two, so the backend required
its own corpus at the granularity it models. This section documents the
pipeline; it is the experimental setup's main methodological contribution and
is fully reproducible from a single script.

\subsection{Pipeline}

\begin{enumerate}
\item \textbf{Pin a compatible revision.} The reference instrumentation patch
  applies cleanly at Asterinas v0.16.0 (matching the toolchain in the
  \texttt{asterinas/asterinas:0.16.0} image), not at \texttt{main}.
\item \textbf{Instrument.} Apply the benchmark's reference patch, which adds
  a trace module to the kernel's spinlock emitting one serial JSON event per
  lock operation, plus a new two-thread kernel test (below).
\item \textbf{Build and run.} Inside the container: build the initramfs and
  run the instrumented kernel test under QEMU (software emulation), collecting
  the event stream from the serial port.
\item \textbf{Parse.} Fold the event stream into
  $(\mathit{pre}, \mathit{action}, \mathit{post})$ NDJSON windows.
\end{enumerate}

\subsection{A two-thread contention scenario without deadlock}

The reference spinlock kernel tests are single-actor: every operation is
attributed to one thread, so contention, the behavior that makes a lock
interesting, never appears in the trace. But a blocking \texttt{lock()}
from a second actor in a single-CPU kernel test would spin forever. The
scenario therefore expresses the second actor's contention through
non-blocking \texttt{try\_lock} calls that fail without blocking:

\begin{lstlisting}[style=code]
actor0.lock()      -> TryAcquireBlocking(0), AcquireSuccess(0)  // 0 holds
actor1.try_lock()  -> AcquireFail(1)          // contention, no deadlock
actor0.release()   -> Release(0)
actor1.lock()      -> TryAcquireBlocking(1), AcquireSuccess(1)  // 1 holds
actor0.try_lock()  -> AcquireFail(0)          // contention
actor1.release()   -> Release(1)
actor0.try_lock()  -> AcquireSuccess(0)       // try from free succeeds
actor0.release()   -> Release(0)
\end{lstlisting}

A second, longer scenario (more alternation of the holder, more failed and
successful \texttt{try\_lock} attempts, and same-thread reacquisition)
extends the corpus from 8 windows to \textbf{28 windows} (17
\texttt{AcquireLock}, 11 \texttt{ReleaseLock}), the corpus used in all
experiments below.

\subsection{Model-independent state windows}

The kernel emits low-level events (\texttt{TryAcquireBlocking},
\texttt{AcquireSuccess}, \texttt{AcquireFail}, \texttt{Release}). The parser
folds these into windows keyed on the \emph{objective} spinlock state,
$\{\mathit{lockHeld}, \mathit{lockHolder}\}$, the real observable ownership,
and relies on the projection rule to ignore any model's internal
bookkeeping (thread status, call type). A failed acquire under contention
produces a window whose post-state ownership is unchanged; a successful
acquire records the new holder together with whether it arrived via the
blocking (\texttt{lock}) or non-blocking (\texttt{try}) path. Keying the
corpus on observable state keeps the ground-truth oracle independent of any
model under test.

\section{Experiment 1: four models, four phases}
\label{sec:models}

\subsection{Setup}

Four Claude models spanning the capability range (Haiku~4.5, Sonnet~4.6,
Opus~4.8, and Fable~5, Anthropic's most capable model at the time of
writing) each generated one \jssam{} specification for \texttt{spin}.
These are the models' production names, not pseudonyms; the exact API
identifiers are \texttt{claude-haiku-4-5}, \texttt{claude-sonnet-4-6},
\texttt{claude-opus-4-8}, and \texttt{claude-fable-5} (accessed May--July
2026), pinned in the repository's \texttt{config/models.yaml}. Each model
generated via a single prompt (\texttt{direct\_call}), and the same artifact was scored in
all four phases. Phase 2 and 4 exploration ran through the
\texttt{sam-pattern} library's bounded checker at the task's default depth
bound of 6; Phase 3 used the 28-window corpus; Phase 4 checked three expert
invariants (MutualExclusion, LockStatusConsistency, NoDeadlock) translated
via the direct API path.

\subsection{Results}

\begin{table}[ht]
\centering
\caption{Four-phase \jssam{} results on \texttt{spin} (28-window corpus),
models as columns. P3 per-action rows give windows passed / windows of that
action. Base rate: 6 of the 28 windows (all contention acquires) have
post $=$ pre, so the identity function scores 21.4\% overall and 6/17 on
acquires; Haiku's column equals that base rate exactly. Conditional on the
22 state-changing windows the P3-overall row reads 86.4 / 36.4 / 36.4 / 0\%.}
\label{tab:models}
\small
\begin{tabular}{lcccc}
\toprule
 & Opus 4.8 & Fable 5 & Sonnet 4.6 & Haiku 4.5 \\
\midrule
P1 syntax          & pass & pass & pass & pass \\
P2 (distinct states) & pass (7) & pass (7) & pass (7) & pass (\textbf{1}, vacuous) \\
P3 overall         & \textbf{89.3\%} (25/28) & \textbf{50.0\%} (14/28) & \textbf{50.0\%} (14/28) & \textbf{21.4\%} (6/28) \\
P3 Acquire         & 14/17 & 14/17 & 14/17 & 6/17 \\
P3 Release         & 11/11 & \textbf{0/11} & 0/11 & 0/11 \\
P4 invariants      & 3/3 & 3/3 & 3/3 & 3/3 \\
\bottomrule
\end{tabular}
\end{table}

Two results stand out.

\paragraph{Only the reality check discriminates.}
Every model passes Phase 1, Phase 2, and Phase 4. On this task those three
phases have \emph{zero} discriminating power. The checker reports an
identical 326{,}592 ``states explored'' per model, but that number is not
evidence of structurally equivalent state spaces: it is the checker's
\emph{step count}, that is, safety-callback invocations over the
intent-permutation tree, $(d{+}1)\cdot 6^{d} = 7\cdot 6^{6}$, a pure
function of the contract-pinned intent domain and depth bound, identical for
any conforming spec and therefore model-independent. Counting \emph{distinct
semantic states} (unique model snapshots) instead separates the specs: 7 for
Opus, Fable, and Sonnet, but \textbf{1 for Haiku}, whose intent actions drop
their arguments so every proposal is rejected, so its exploration never leaves
the initial state. Haiku's Phase-2 pass is thus \emph{vacuous} (no crash,
deterministic, and serializable hold trivially for a spec that cannot move),
and its Phase-4 invariants are satisfied over that single state. Phase 2
verifies checkability, not semantic adequacy; the distinct-state count, now
reported by the checker, is the semantic signal. Phase 3 spreads the models
across a $4\times$ range, and, restricted to comparisons it can make,
orders Opus above the mid tier above Haiku. A base-rate audit sharpens the
spread: 6 of the 28 windows (all contention acquires)
have post $=$ pre, so the \emph{identity function} scores 21.4\% overall and
35.3\% on acquires. Haiku's column is exactly that base rate: its
``Acquire 6/17'' is precisely the six freebies, and it passes \emph{zero}
windows requiring a state change, so conditional on the 22 change windows
the spread is not 21.4--89.3\% but \textbf{0--86.4\%} (Opus 19/22, Fable and
Sonnet 8/22, Haiku 0/22). Every arm of every model collects all six
freebies; the conditional-on-change rate is the discriminating metric and
accompanies headline rates hereafter. This independently reproduces, in
a non-formal language, the central SysMoBench finding for \tlaplus{}: internal
consistency is cheap; conformance with the real system is where specifications
fail~\cite{sigops2026}.

\paragraph{General capability does not predict modeling accuracy.}
The most striking column is Fable~5, the most capable model tested, scoring
50\%, below Opus~4.8 and tied with mid-tier Sonnet. The cause is specific
and systematic, not noise: Fable's specification handles acquisition
competently (14/17) but \emph{no-ops every single release} (0/11). After any
\texttt{ReleaseLock}, its model leaves the lock held:

\begin{lstlisting}[style=code]
ReleaseLock  expected {lockHeld: false, lockHolder: null}
             got      {lockHeld: true,  lockHolder: 0}
\end{lstlisting}

The release logic \emph{reads} correctly in isolation (it guards on the
holder, then clears the state) but on replay the release never takes
effect: a categorical, whole-action modeling defect. Crucially, this
specification passed syntax, completed the full bounded exploration (7
distinct states) without violation, and satisfied all three invariants,
including mutual exclusion. (Its release \emph{does} fire when reached from
the initial state, where its auxiliary \texttt{threadStatus} guard holds; the
defect is that replay pins only the observable pre-state, leaving the guard
unsatisfiable, so exploration from init cannot expose it.) Only replay
against the real kernel's transitions caught it.

The recurring failure among the passing acquisitions is also precise: all
three 14/17 models fail exactly the three \texttt{try\_lock}-from-free
windows, the non-blocking acquire that should succeed on a free lock.
Blocking acquisition (abundant in training data and in textbook treatments)
is modeled correctly by every model except Haiku; the less common
non-blocking path is where they slip. This echoes, at micro scale,
SysMoBench's ``textbook modeling'' diagnosis: models reproduce the familiar
template and miss the implementation-specific path.
\section{Experiment 2: repair from trace feedback}
\label{sec:repair}

\subsection{Setup}

If a model is shown \emph{exactly} which transitions its specification got
wrong (the pre-state, the action and its data, the real system's
post-state, and what the model produced instead), can it fix the
specification? For each model we ran one repair round: score the original
specification on Phase 3; build a repair prompt containing the full original
module plus every failing window; ask the model to rewrite the module;
confirm the result still passes Phase 1; re-score Phase 3 on the same
28-window corpus.

\subsection{Results}

\begin{table}[ht]
\centering
\caption{One-round repair from Phase-3 feedback (same 28-window corpus;
original run; the generalization-audit replication differs for Sonnet,
which repaired to 28/28 there). Models as columns.}
\label{tab:repair}
\small
\begin{tabular}{lcccc}
\toprule
 & Opus 4.8 & Fable 5 & Sonnet 4.6 & Haiku 4.5 \\
\midrule
Baseline P3      & 89.3\% (25/28) & 50.0\% (14/28) & 50.0\% (14/28) & 21.4\% (6/28) \\
Repaired P3      & \textbf{100\%} (28/28) & \textbf{100\%} (28/28) & 60.7\% (17/28) & 21.4\% (6/28) \\
Acquire (base$\to$rep) & 82\% $\to$ 100\% & 82\% $\to$ 100\% & 82\% $\to$ 100\% & 35\% $\to$ 35\% \\
Release (base$\to$rep) & 100\% $\to$ 100\% & \textbf{0\% $\to$ 100\%} & 0\% $\to$ 0\% & 0\% $\to$ 0\% \\
\bottomrule
\end{tabular}
\end{table}

\paragraph{Self-correction separates the models, but the middle tier is noise.}
In the run of Table~\ref{tab:repair}, Opus and Fable repair to a perfect
28/28 in a single round, Sonnet repairs \emph{partially} (it fixes the
\texttt{try\_lock} acquisition defect but not its release path), and Haiku
does not improve at all: superficially a clean three-tier gradient
(full / partial / none). A replication run (part of the generalization
audit below, with artifacts saved) revised this: \emph{Sonnet also repaired
to 28/28}. The boundary that is stable across both runs is two-tier:
Opus, Fable, and (unstably) Sonnet can exploit transition-level feedback;
Haiku cannot; and single-round repair outcomes for mid-tier models are
sampling-dependent at $N{=}1$.

\paragraph{The Fable reversal.}
Fable's baseline looked no better than Sonnet's, and its failure mode,
a whole action silently broken, looked worse. Yet under repair, Fable
fully recovers: its release defect was a \emph{recoverable slip}, not a
capability ceiling. One-shot specification accuracy and
repair-from-feedback are different axes of capability, and a benchmark that
measures only the former will misrank models for any workflow that includes
a correction loop, which realistic agentic formal-modeling workflows
do~\cite{sigops2026}.

\paragraph{Generalization audit: the repairs are fixes, not memorization.}
Repaired specifications were originally re-checked only on Phases 1 and 3,
on the same 28 windows quoted in the repair prompt, leaving open that a
repair might \emph{memorize} the failing windows (branch on the pre-state,
return the expected post), which would make this an experiment about
prompt-following rather than modeling. An audit re-ran the repair loop with
artifacts saved and scored each baseline and repaired specification on the
seen corpus, on \emph{held-out} windows, and on Phases 2 and 4. The 28 seen
windows collapse to 8 distinct (pre-state, action, data) combinations; the
held-out set is the 10 combinations of the observable domain the corpus
never shows (acquire on a held lock, release by a non-holder, release of a
free lock). All repaired specifications pass all held-out windows, reach
the same 7 distinct semantic states as a correct specification under
bounded exploration (Haiku's unrepaired specification remains at 1), and
hold all invariants. Direct inspection found no memorization (no
state-equality tables or window literals); the successful repairs are the
same root-cause fix, moving the release ownership check from the auxiliary
\texttt{threadStatus} guard (unsatisfiable under a pinned observable
pre-state) to the authoritative \texttt{lockHeld}/\texttt{lockHolder},
with Sonnet's repair documenting the diagnosis in a comment. One
structural honesty note: because the kernel instrumentation emits acquire
events at acquisition success, every held-out combination is an observable
no-op, so held-out scoring alone refutes only memorizers with unsafe
defaults; the inspection and Phase-2/4 evidence carry the conclusion,
and a state-changing held-out set (as \texttt{locksvc} affords) is the
right vehicle for a stronger version of this audit.

\paragraph{What this does and does not measure.}
The repair prompt contains the failing windows themselves, so the experiment
measures ``can the model exploit precise, correct feedback,'' not blind
improvement. That is the intended question (it is exactly the signal
available inside an agentic repair loop) but it is not a from-scratch
re-test. A multi-round follow-up
(\texttt{scripts/repair\_multiround.py}) answers the convergence question:
Haiku does \emph{not} converge (three rounds of full-failure feedback
leave it at 6/28 with its exploration still confined to one state,
6$\to$6$\to$6, with every round's spec saved and audited), while Sonnet
full-repairs in round one again (its second full repair in three runs,
further evidence the original ``partial'' tier was a sampling artifact).

\section{Experiment 3: \jssam{} versus \tlaplus{}}
\label{sec:vstla}

\subsection{Setup}

The head-to-head the backend exists to inform: the same four models, the same
\texttt{spin} task, one \texttt{direct\_call} generation per model per
language, scored on Phase 1 (syntax: SANY vs.\ module validation), Phase 2
(model checking: TLC vs.\ bounded exploration), and Phase 4 (invariant
verification, direct translator for both languages). Phase 3 is deliberately
\emph{not} compared: \jssam{}'s Phase 3 is a self-contained direct replay
while \tlaplus{}'s is a heavyweight coding-agent path, that is, different
mechanisms rather than a fair one-to-one. Phase 4 is compared as pass/fail
because the two languages ship different invariant-template counts for
\texttt{spin} (seven for \tlaplus{}, three for \jssam{}), and the \jssam{}
three are safety-only, a qualitative weakness examined in
\S\ref{sec:limitations}: as implemented, \jssam{} Phase 4 could not have
failed the never-releasing defect of \S\ref{sec:models} in principle.

\subsection{Results}

Table~\ref{tab:vstla} summarizes the per-phase outcome and Table~\ref{tab:tladetail} the per-model \tlaplus{} detail.

\begin{table}[ht]
\centering
\caption{\jssam{} vs.\ \tlaplus{} on \texttt{spin}: models producing a passing artifact per phase.}
\label{tab:vstla}
\small
\begin{tabular}{lccc}
\toprule
 & P1 syntax & P2 model-checkable & P4 invariants \\
\midrule
\jssam{}  & 4/4 & \textbf{4/4} & \textbf{4/4} \\
\tlaplus{} & 4/4 & \textbf{1/4} & \textbf{1/4} \\
\bottomrule
\end{tabular}
\end{table}

\begin{table}[ht]
\centering
\caption{Per-model \tlaplus{} detail.}
\label{tab:tladetail}
\small
\begin{tabular}{lcccp{4.6cm}}
\toprule
Model & P1 & P2 & P4 & P2 failure cause \\
\midrule
Claude Opus 4.8   & pass & \textbf{fail} & fail & unbounded \texttt{CHOOSE} (TLC rejects) \\
Claude Fable 5    & pass & pass (159 distinct states) & pass (7 inv.) & (none) \\
Claude Sonnet 4.6 & pass & \textbf{fail} & fail & unbounded \texttt{CHOOSE} (TLC rejects) \\
Claude Haiku 4.5  & pass & \textbf{fail} & fail & config generation fell back; no usable \texttt{.cfg} \\
\bottomrule
\end{tabular}
\end{table}

\paragraph{Syntax is not where the languages differ.}
Every model writes syntactically valid specifications in \emph{both}
languages, consistent with SysMoBench's finding that frontier models have
essentially saturated \tlaplus{} syntax~\cite{sigops2026}. The divergence is
at Phase 2: whether the specification is actually \emph{checkable}. All four
\jssam{} modules run; one of four \tlaplus{} specifications does. Each
\tlaplus{} failure is attributable to a specific structural cause.

\paragraph{Cause 1: a \tlaplus{} semantic pitfall JavaScript does not have.}
Opus and Sonnet both modeled ``no lock owner'' as
\texttt{NULL == CHOOSE v : v \textbackslash notin Threads}. SANY accepts
this (syntax passes) but TLC cannot enumerate a \texttt{CHOOSE} whose
bound variable ranges over an unbounded domain (\texttt{CHOOSE} itself is
deterministic), so model checking fails. The idiom is natural, looks
correct, and is a well-known TLC trap, though the indictment is of LLM
\tlaplus{} fluency rather than of the language: no experienced \tlaplus{}
author models absence this way, and agentic workflows already write around
it in practice~\cite{specula}. JavaScript has a native \texttt{null},
so the corresponding \jssam{} specifications express the same concept with no
trap available to fall into. This structural hazard caught two of four
models, \emph{including the model with the best \jssam{} conformance score}
(Opus). It is a clean instance of the hypothesis's mechanism: the failure is
not ``the model cannot model a spinlock'' (its JavaScript model of the same
system scored 89.3\% against real traces) but ``the formal language exposes a
semantic surface on which competent models slip.''

\paragraph{Cause 2: a configuration surface \jssam{} does not have.}
\tlaplus{} model checking requires a separate \texttt{.cfg} artifact
(constants, init/next, invariants) that the harness must generate to match
the specification; for Haiku, that generation fell back and produced no
usable configuration (zero states explored). A \jssam{} module is
self-contained and executable: there is no second artifact to generate,
so this entire failure mode is structurally absent. We flag the attribution
honestly: the \texttt{CHOOSE} failures belong to the models; the Haiku
failure belongs to the harness. But the asymmetry itself is structural:
one language's evaluability depends on a fragile secondary artifact and the
other's does not.

\paragraph{Net.}
On this task, models reach an end-to-end evaluable specification far more
readily in \jssam{} (4/4 through Phase 4) than in \tlaplus{} (1/4). Read
narrowly, this is evidence for the hypothesis in its practical form:
\emph{host-language familiarity plus a self-contained executable artifact
lowers the barrier to a usable specification.} It does not, by itself, show
that \jssam{} specifications are more \emph{faithful}; it shows they are
far more often checkable at all, which is the precondition for faithfulness
to be measured. And given the per-cause attribution above and the guardrail
result of \S\ref{sec:crosslang}, we treat the 4/4-vs-1/4 figure throughout
this paper as a within-experiment observation about default-prompt
robustness, not a conclusion about the languages, and as a \emph{lower
bound} on \tlaplus{}: a one-line prompt guardrail removes the gap entirely
(\S\ref{sec:crosslang}), and deployed agentic workflows already write
around the \texttt{CHOOSE} idiom~\cite{specula}. One asymmetry in this
experiment must be named at the point of claim rather than in the
limitations: the 1/4 figure is a one-shot number for a language whose
failures we have just argued are trivially repairable, whereas the \jssam{}
side receives a repair round in Experiment 2. Symmetry, together with Erik Meijer's suggestion in review that a fair comparison must steer and repair the larger-surface-area language on the same terms~\cite{meijer2026}, demands the same for \tlaplus{}: one repair round per failing specification (the model's own
specification, its configuration, and TLC's verbatim error output, nothing
else) repairs \emph{both}. Sonnet's rewrite adopts the string sentinel
the guardrail prescribes (\texttt{NULL == "NULL"}; TLC passes, 111
distinct states); Opus keeps the \texttt{CHOOSE} but bounds it over a
singleton set (TLC passes, 159 distinct states); both preserve the action
structure. Carried through the invariant pipeline, both repaired
specifications hold all seven expert invariants (7/7 each, matching the
one \tlaplus{} specification that passed unrepaired), after resolving
one further harness artifact the rescore itself exposed: the pipeline
appends each translated invariant under its template name, so a
specification that defines its own \texttt{MutualExclusion}, as Sonnet's
repair does, draws a duplicate-definition parse error that scores as an
invariant failure. Per this paper's per-cause discipline we report that
as a benchmark defect, not a specification one. The 1/4 figure is thus
bounded from two directions: a one-line
guardrail prevents the failure, and one round of TLC feedback repairs it
fully, consistent with the repair-axis findings of Experiments 2 and 5,
and further evidence that the evaluability gap is shallow.
Experiment 4 (\S\ref{sec:crosslang})
measures faithfulness directly, under a design built to make the two
languages comparable, and resolves it in an unexpected place.

\subsection{The deployed agent-mediated path}
\label{sec:agentpath}

It is worth asking what the benchmark's \emph{as-deployed} \tlaplus{} Phase 3
(a coding agent that instruments the system, captures traces, cuts windows,
and drives TLC per window) makes of these same four specifications.
We ran it: each Experiment-3 specification through the deployed launcher,
with a fresh QEMU kernel capture per run and the scope pinned to the two target actions (Table~\ref{tab:agentpath}).

\begin{table}[ht]
\centering
\caption{The Experiment-3 \tlaplus{} specifications under the deployed
agent-mediated Phase 3, beside the controlled result. The two paths do not
score the same window set (see text).}
\label{tab:agentpath}
\small
\begin{tabular}{llp{4.2cm}p{2.6cm}}
\toprule
Spec & Controlled (Exp.\ 3) & Deployed agent path & Cause \\
\midrule
Fable 5   & TLC pass & \textbf{22/22} state-changing windows & (none) \\
Opus 4.8  & TLC fail & \textbf{unevaluable} (TLC error on all windows) & unbounded \texttt{CHOOSE} \\
Sonnet 4.6 & TLC fail & \textbf{unevaluable} (identical) & unbounded \texttt{CHOOSE} \\
Haiku 4.5 & harness fault & \textbf{19/19} on its own action semantics & agent supplies the config \\
\bottomrule
\end{tabular}
\end{table}

Three observations follow, each modest but worth stating precisely.

First, the per-cause attribution of Experiment 3 survives the ecological
test. The harness configuration fault does not exist on the deployed path
(the agent writes its own configurations and checking modules), so
Haiku's specification becomes evaluable, confirming that failure belonged
to our harness. The \texttt{CHOOSE} idiom persists: the agent reproduced
the failure against the pristine specification and its own original
configuration, and correctly concluded that no checking-side workaround
exists, since the sentinel is a derived operator rather than a rebindable
constant. Deployed evaluability is therefore 2/4, against 1/4 controlled;
the deployed path is the more forgiving instrument, and what remains is
exactly the model-written idiom.

Second, the two paths do not measure the same thing. The deployed agent
excluded the six contention no-op windows as out of scope (documented
exclusions, with precedent in the benchmark's reference materials), where
the controlled oracle requires them as identity transitions. Deployed and
controlled pass rates are computed over different window sets and should
not be compared numerically.

Third, and most consequential for interpreting agent-mediated scores,
the deployed oracle is \emph{interpretive}. Haiku's specification models
\texttt{AcquireLock} as request initiation (idle to trying, ownership
unchanged) rather than as acquisition success. The agent read the
specification, cut its windows on the corresponding trace events, and
scored it 19/19, a perfect score on the specification's own reading of
the action names, for the same model whose \jssam{} specification scored
the identity base rate against the pinned-schema oracle. Neither score is
wrong; they answer different questions. The pinned oracle asks ``does the
specification reproduce the agreed observable semantics''; the agent asks
``is the specification consistent with the traces under its own
vocabulary.'' The gap between those questions is invisible until one runs
both instruments on the same artifact, which is the practical argument for
keeping a controlled conformance mode alongside the deployed one
(\S\ref{sec:scope}).
\section{Experiment 4: cross-language conformance under a controlled contract}
\label{sec:crosslang}

Experiment 3 could not compare Phase 3 across languages because the two
backends implement it by incomparable mechanisms: \jssam{} runs a mechanical
replay, while \tlaplus{} runs a coding-agent path, necessary because a
free-form \tlaplus{} specification invents its own variable names, so mapping
trace states onto spec states requires interpretation. A naive comparison
would measure the replay machinery, not the specifications. This experiment
removes that asymmetry with a pre-registered, paired design; the full
methodology note, written to be socialized with the benchmark maintainers, is
included in the artifact repository.

\subsection{Design}

\paragraph{Equalize the observable-state contract.}
The degree of freedom that forces \tlaplus{} onto the agent path, namely free
choice of variables, is one the \jssam{} prompt already removes by pinning
\texttt{getState()}'s shape. We remove it symmetrically: a constrained
\tlaplus{} prompt mandates \texttt{VARIABLES lockHeld, lockHolder} (the trace
schema), action operators \texttt{AcquireLock(thread, callType)} /
\texttt{ReleaseLock(thread)}, and a TLC-safe sentinel
(\texttt{NONE == "none"}, not \texttt{CHOOSE}). Both languages then answer
the identical question, namely \emph{express this system over this observable
state}, and trace states map onto spec states with no interpretation.

\paragraph{A direct TLC replay path (no agent), functional on both sides.}
With the contract fixed, \tlaplus{} transition validation becomes mechanical,
mirroring \jssam{}'s \texttt{setState} $\to$ action $\to$ diff: for each
window, a synthesized module pins the pre-state in its initial predicate and
takes exactly one step of the window's action. Mere reachability of the
traced post-state would be too weak a criterion: a \tlaplus{} action is a
\emph{relation}, and an over-permissive action can reach the correct
post-state while also admitting wrong ones, passing an existential check
that the functional JS replay (one \texttt{next} state, compared exactly)
would fail. The replay therefore enforces the same functional relation on
both languages: it enumerates the action's one-step image from the pinned
pre-state and requires that the image be exactly the traced post-state
(branching factor 1; over-permissiveness fails the window). The check was
validated in three directions: a contract-following specification passes
28/28 (positive control); the same specification with \texttt{ReleaseLock}
mutated to a no-op fails exactly the 11 release windows (negative control);
and a deliberately permissive specification (holder set to \emph{any}
thread on acquire) passes 28/28 existentially but only 11/28
functionally, with branching factor 2 (permissive control), demonstrating
that the functional check is strictly stronger and necessary. An audit of
all 20 generated \tlaplus{} specifications found every one deterministic:
existential and functional scores coincide on every window, and the maximum
branching factor is 1. The \tlaplus{} results below are therefore earned
under a relation as strict as the JS one, not inflated by existential
slack.

Two commitments of this oracle must be named, because they are
modeling-paradigm choices rather than neutral ground. First, requiring
every trace event to map onto a named, \emph{total}, successor-producing
action is the \barenext{}/JS paradigm by construction. Idiomatic \tlaplus{}
specifies \emph{enabling conditions} and relies on stuttering
($[\mathit{Next}]_{\mathit{vars}}$ always admits
$\mathit{vars}' = \mathit{vars}$): a failed acquisition is naturally a
disabled action plus a stutter, which this oracle scores as
\emph{unscoreable} rather than as a legitimate no-op, penalizing
enabling-condition specifications independent of their correctness.
Wherever the cross-language comparison is drawn, we therefore report the
conditional (paradigm-fair) score alongside the unconditional one, and we
flag a stutter-aware oracle (a $post = pre$ window satisfiable by a
stuttering step) as the fairer instrument for such specifications.
Second, the functional branching-factor-1 requirement encodes a
determinism assumption that holds here only because the trace scenarios
are deterministic by construction; for genuinely concurrent systems a
relational next-state is fidelity, not over-permissiveness, and the oracle
must weaken to set-conformance (traced post-state in the one-step image,
with tightness reported separately) before this methodology reaches
Raft-class protocols.

\paragraph{Four arms, because the prompt and the contract are both confounds.}
An initial two-arm run (deployed \jssam{} prompt vs.\ constrained \tlaplus{})
produced what looked like a decisive \tlaplus{} reversal. But the constrained
\tlaplus{} prompt \emph{states the exact single-step observable semantics}
(free $\to$ acquire; held $\to$ unchanged; holder $\to$ release), which the
deployed \jssam{} prompt does not: a difference in prompt content, not in
language. And even with prompts equalized, the two substrates still differ in
a second way: the SAM contract obliges an executable module with
proposal/acceptor wiring and auxiliary state, while the \tlaplus{}
specification is a bare two-variable relation. We therefore ran four arms
initially:
\textbf{JS(deployed)}, the shipped \jssam{} prompt;
\textbf{JS(constrained)}, the deployed prompt plus the \emph{identical}
semantics block used by the \tlaplus{} arm;
\textbf{plain-JS}, the same constrained semantics, but the specification
is a bare \texttt{next(state, action, model)} transition function
(argument names follow SAM usage) over only
$\{\mathit{lockHeld}, \mathit{lockHolder}\}$: JavaScript, yet structurally
isomorphic to the \tlaplus{} relation, with no SAM library and no auxiliary
state; and \textbf{TLA(constrained)}. JS(constrained) vs.\ TLA(constrained)
varies language and contract together; plain-JS vs.\ TLA(constrained) varies
\emph{only} the language; JS(constrained) vs.\ plain-JS varies \emph{only}
the contract.

\paragraph{Completing the factorial: the missing bare-next-without-semantics cell.}
These arms form an \emph{incomplete} $2{\times}2$ in
(contract, prompt): the \barenext{} arm always carried the semantics block, so
``contract minimality buys transcription fidelity'' was established only
conditional on spec-level semantics in the prompt, and any
prompt-vs-contract effect ordering rested on the incomplete design. We added
the missing cell, \textbf{plain-JS(no-semantics)}: the identical \barenext{}
contract (module shape, two observable keys, action names, data schema, and
a description of the replay \emph{mechanism}) with the entire single-step
semantics block removed, so the model must derive the transition semantics
from the Rust source alone. A mirrored \tlaplus{} derivation arm (the
constrained \tlaplus{} contract, semantics block likewise removed) was added
alongside it, completing the $3{\times}2$ of Table~\ref{tab:crosslang}; its
result, the study's only cross-language split, is analyzed in the
findings below, with the full audit trail preserved in
\S\ref{sec:limitations}.

\paragraph{Replication, pairing, and checkability.}
$N{=}5$ generations per model per arm (80 in total), all scored per-window on
the same 28-window kernel corpus. A pooled per-window McNemar test would be
\emph{pseudo-replicated} here: generations collapse to 1--2 unique behavioral
fingerprints per (model, arm), so pooling $5\times28$ windows counts the same
underlying defect up to five times and inflates any $\chi^2$. The analysis of
record is therefore at the generation level: we report per-arm uniqueness
counts and an exact two-sided permutation test with the \emph{generation} as
the unit (statistic: difference in mean per-generation pass rate; all
$\binom{10}{5}=252$ relabelings; the smallest attainable $p$ is
$2/252\approx0.0079$, so starred results sit at the test's floor). Because
the generations collapse to so few distinct behaviors, we present $\Delta$
as the primary quantity and read the permutation $p$-values as descriptive
rather than as hypothesis tests: the effect sizes are large and uniform;
the inferential machinery around them is at its resolution floor. The
pre-registered checkability policy (conditional vs.\ unconditional scoring)
proved moot in the constrained arms: \emph{zero} windows were unscoreable
in any constrained or plain-JS generation. (It becomes load-bearing in the
derivation arms below.)

\subsection{Results}

\begin{table}[ht]
\centering
\caption{Cross-language Phase-3 conformance (mean unconditional pass rate
over $N{=}5$ generations per cell; 28-window corpus), models as columns. The
rows form a full $3{\times}2$ factorial: contract
(SAM / \barenext{} / \tlaplus{}) $\times$ prompt (without / with the
semantics block). All shortfall in the \tlaplus{}-no-semantics row is
\emph{unscoreable} (guard-idiom partiality), not wrong post-states;
conditional rates there are 100\%.}
\label{tab:crosslang}
\small
\begin{tabular}{lcccc}
\toprule
Arm & Opus 4.8 & Fable 5 & Sonnet 4.6 & Haiku 4.5 \\
\midrule
SAM, no sem.\        & 81.4\% & 57.9\% & 50.0\% & 28.6\% \\
SAM, sem.\           & 89.3\% & 95.7\% & \textbf{100\%} & 40.7\% \\
\barenext{}, no sem.\ & \textbf{100\%} & \textbf{100\%} & \textbf{100\%} & \textbf{100\%} \\
\barenext{}, sem.\   & \textbf{100\%} & \textbf{100\%} & \textbf{100\%} & \textbf{100\%} \\
\tlaplus{}, no sem.\ & 78.6\% & \textbf{100\%} & 78.6\% & 87.1\% \\
\tlaplus{}, sem.\    & \textbf{100\%} & \textbf{100\%} & \textbf{100\%} & \textbf{100\%} \\
\bottomrule
\end{tabular}
\end{table}

\begin{table}[ht]
\centering
\caption{Generation-level analysis (a pooled per-window McNemar would be
pseudo-replicated here and is not used). ``Unique'' is the count of distinct
behavioral fingerprints
(per-window outcome vectors) among the 5 generations of each arm, ordered
deployed-JS / constrained-JS / plain-JS / \tlaplus{}. $\Delta$ is the
difference in mean per-generation pass rate vs.\ \tlaplus{}; $p$ from the
exact permutation test. No significance markers are used: with $N{=}5$
paired generations the smallest attainable $p$ is $2/252\approx0.0079$, a
resolution floor at which $p$-values cannot rank effects, so read $\Delta$ as
the primary quantity and $p$ descriptively. The plain-JS arm is omitted: it
ties \tlaplus{} in every generation ($\Delta{=}0$, $p{=}1$).}
\label{tab:mcnemar}
\small
\begin{tabular}{lccc}
\toprule
Model & Unique (dep/con/plain/\tlaplus{}) & JS(deployed) vs.\ TLA & JS(constrained) vs.\ TLA \\
\midrule
Opus 4.8   & 2/1/1/1 & $\Delta{=}.186$, $p{=}.0079$ & $\Delta{=}.107$, $p{=}.0079$ \\
Fable 5    & 2/2/1/1 & $\Delta{=}.421$, $p{=}.0079$ & $\Delta{=}.043$, $p{=}.44$ \\
Sonnet 4.6 & 1/1/1/1 & $\Delta{=}.500$, $p{=}.0079$ & $\Delta{=}0$, $p{=}1$ \\
Haiku 4.5  & 1/2/1/1 & $\Delta{=}.714$, $p{=}.0079$ & $\Delta{=}.593$, $p{=}.0079$ \\
\bottomrule
\end{tabular}
\end{table}

\subsection{Findings}

\paragraph{Prompt prescriptiveness is a large, real confound that the
completed factorial nonetheless demotes from the headline.}
Adding the same semantics block to the JavaScript prompt, changing nothing
else, moved Fable from 57.9\% to 95.7\% and Sonnet from 50.0\% to a
perfect 100\%. The two-arm run alone would suggest a decisive language
reversal, read as ``most of that effect was the prompt.'' The completed
factorial shows that reading is an artifact of examining only the SAM
column: at the generation level the
within-SAM prompt effect is large for exactly those two models (Fable
$\Delta{=}.379$, $p{=}.024$; Sonnet $\Delta{=}.500$, $p{=}.0079$) and
absent for
Opus ($\Delta{=}.079$, $p{=}1$) or Haiku ($\Delta{=}.121$, $p{=}1$), whereas
the \emph{contract} effect without any semantics block (SAM vs.\ \barenext{},
both promptless) is large and uniform for \emph{all four} models
($\Delta{=}.186/.421/.500/.714$, each at the permutation floor). Prompt
prescriptiveness remains a result in its own right, a first-class
experimental variable large enough to masquerade as a language effect, but
on this task it is the model-dependent factor, and the contract is the
uniform one.

\paragraph{The missing cell: the \barenext{} contract needs no semantics block.}
The \barenext{}-without-semantics arm reaches \textbf{100\% for every
generation of every model}, including Haiku, at 28.6\% under the SAM contract
with the same absent semantics. All 20 generations are textually distinct
(5/5 unique per model, saved with the study artifacts) yet behaviorally
identical and correct: given a bare \texttt{next()} target, every model
derives the observable single-step semantics from the kernel source alone.
This settles the conditionality the incomplete design left open: contract
minimality buys transcription fidelity \emph{unconditionally} on spec-level
semantics in the prompt. Two honesty notes travel with this. First, with both
\barenext{} cells and \tlaplus{} at 100\%, effect \emph{ordering} at the top
is not identifiable: the within-\barenext{} prompt effect and any
prompt$\times$contract interaction have no headroom to appear (a ceiling
effect); what is identifiable is that contract $\geq$ prompt for every model,
strictly greater for Opus and Haiku, and that the \barenext{} contract
suffices without semantics while the semantics block within SAM does not
(three of four models below 100\%). Second, the 100\% ceiling itself is a
property of this task's tiny observable relation; the \texttt{locksvc}
replication (\S\ref{sec:locksvc}) shows the \barenext{} result generalizing
to a richer state, but the factorial has not yet been re-run there.

\paragraph{The derivation arms split the languages, and the split is the
oracle's paradigm, not the language.}
With the semantics block removed from both minimal contracts, \barenext{} JS
reaches 100\% for every model while \tlaplus{} falls to 78.6--100\%
($\Delta{=}.214$ for Opus and Sonnet, at the permutation floor;
Table~\ref{tab:crosslang}). The mechanism contains no wrong post-states:
every \tlaplus{} shortfall window is \emph{unscoreable}, because unguided
models write the idiomatic guarded action (enabled only when the lock is
free), and under this oracle a disabled action on a pinned contention
pre-state has no successor. The \barenext{} contract's \emph{totality} forbids
that idiom structurally; \tlaplus{} permits it; and the semantics block's
``held $\Rightarrow$ unchanged'' clause is precisely what converts the
guard into a no-op branch. Scored conditionally (the paradigm-fair
reading, per the oracle commitments stated in the design), the
guard-idiom specifications are 100\% correct on every window they can
replay, and the languages tie here too. ``Contract first, prompt second''
therefore refines to: \emph{the prompt matters exactly where the contract
under-constrains}, and the unconditional split is a fact about the
oracle's paradigm, reported as headline only because the benchmark's
deployed scoring is unconditional.

\paragraph{A real, one-directional residual survives the prompt control.}
With identical semantics in both prompts, \tlaplus{} reaches 100\%
conformance for \emph{every model and every generation}, while constrained
\jssam{} does so only for Sonnet. At the generation level the residual is
large for Opus and Haiku (both at the permutation floor,
$p{=}.0079$, read descriptively), vanishes for Sonnet, and for Fable does
not replicate
($\Delta{=}.043$, $p{=}.44$): the pooled McNemar had starred Fable's
deficit, and that specific claim was pseudo-replication, now withdrawn. The
residual is largest for the weakest model (Haiku: 40.7\% vs.\ 100\%), and
the direction is uniform: descriptively, no \jssam{} specification ever
passed a window that its \tlaplus{} counterpart failed, in any arm. (When
one arm is at 100\% this is partly structural, as \S\ref{sec:limitations}
notes, but the direction is uniform in the deployed arm as well, where the
gap sits at the permutation floor for all four models.)

\paragraph{The plain-JS arm locates the residual: it is the contract, not
the language.}
The fourth arm ties \tlaplus{} at 100\% for every model and every generation,
with identical per-window outcomes in every paired generation-set
($\Delta{=}0$; no test needed). Haiku is the cleanest single datum:
the weakest model, which reaches only 40.7\% through SAM's machinery,
expresses the identical semantics perfectly as a bare transition function,
in the same language. Three attributions fall out.
\emph{It is not the language}: same JavaScript, different shape, 100\%.
\emph{It is not formality}: \tlaplus{} and plain-JS share the same shape,
a minimal declarative single-step transition over exactly the observable
state, and both sit at 100\%; what the three-arm data suggested was
``formal directness'' is really \emph{minimal declarative shape}.
\emph{It is the SAM contract's defect surface}: proposal/acceptor wiring,
intent firing, and the mandatory auxiliary state
(\texttt{threadStatus}, \texttt{callType}), precisely where
Experiment 1's \texttt{try\_lock}-from-free and release bugs lived. Every
moving part the contract obliges is transcription surface, and the tax falls
hardest on the weakest model (Haiku pays 59 points; Sonnet pays nothing).

\paragraph{Verdict against the pre-registered outcomes.}
Not the strong ``\jssam{} wins'' (the direction is reversed), and not a tie
at the backend level, but at the \emph{language} level a tie is exactly
what the controlled comparison shows: JavaScript expresses the transition
relation as reliably as \tlaplus{} once the specification shape is matched.
The pre-registered outcome list did not anticipate this resolution. For
transparency, the plain-JS arm was designed after the three-arm results were
known, though it was scored blind, by the same mechanical replay, on the
same corpus.

\paragraph{Design implication for \jssam{}.}
The Phase-3 fidelity gap is the cost of the SAM specification contract, not
a language limit, and the two can be decoupled. The backend can offer a
\barenext{} specification shape (\texttt{next(state, action, model)}, the
JavaScript analogue of the \tlaplus{} relation), or score conformance against
a plain transition core over only the observable variables, dropping the
mandatory auxiliary state from the replay contract. SAM's ceremony is not
gratuitous (the instance, intents, and value domains are exactly what feed the
Phase-2/4 bounded explorer) but it should be paid where it buys
something: explorable structure for model checking, not conformance replay.
A prototype confirms the decoupling is practical: a $\sim$40-line generic
breadth-first explorer over \texttt{\{init, next\}} runs the entire pipeline
in the same locked-down sandbox, so the SAM machinery is not \emph{required}
by any phase. Two points keep this precise. First, the SAM checker itself
was never the execution problem: it ran Phases 2 and 4 for every \jssam{} arm
in this paper (the full depth-6 intent exploration each time) without failing
on any specification it was given; the tax is on the \emph{authoring} side,
namely the instance, intents, acceptors, and auxiliary state the model must
hand-write to make a specification checker-consumable. (``Ran without
failing'' is itself a bounded claim: the metric audit in \S\ref{sec:models}
shows the checker also ran to completion on a specification whose exploration
never leaves its initial state, so a clean Phase-2 run certifies
checkability, not adequacy.) Second, nothing is lost at this scale by
the replacement: the SAM checker and the generic explorer are both bounded
safety explorers (reachable states, invariants, deadlock), equivalent in
checking power on these state spaces. The hybrid path keeps the library
checker unchanged (the model writes only the pure \texttt{next}, the
harness generates the scaffolding the checker consumes), relocating the
checker from the specification contract to the tooling, which is where TLC
has always lived on the \tlaplus{} side: no one asks the model to write
TLC. A follow-up study closes the loop with model-generated
specifications: each of the four models generated five \barenext{}
specifications ($N{=}5$, 20 in total, each saved and evaluated as a single
artifact across all phases), and \textbf{20 of 20 pass every phase}
(loadable, Phase-2 clean, Phase-3 28/28, all three Phase-4 invariants
holding). Two caveats travel with this result: \texttt{spin}'s observable
state is tiny (three reachable states), so the \barenext{} explorer's
adequacy must be re-validated on richer tasks; and the 20 generations
collapse to roughly nine unique solutions (the models converge on the same
correct \texttt{next}), which is expected for so small a system but limits
the effective sample. Both caveats are resolved by a second task
(\S\ref{sec:locksvc}).

\paragraph{Checkability, revisited.}
Zero unscoreable generations in the constrained \tlaplus{} arm confirms
Experiment 3's attribution: the earlier ``1/4 model-checkable'' was entirely
the unbounded-\texttt{CHOOSE} trap plus the configuration surface, both of
which the pinned contract and direct replay remove. The checkability gap is
real but shallow, a matter of guardrails, not capability.

\paragraph{A free variance check on Experiment 1.}
The JS(deployed) arm replicates Experiment 1's setting at $N{=}5$. The
single-generation scores (89.3, 50.0, 50.0, 21.4) sit near the five-run means
(81.4, 57.9, 50.0, 28.6), and the model ordering is preserved (with Fable's
mean now slightly above Sonnet's). Experiment 1's qualitative conclusions
survive replication; its exact figures move by up to $\sim$8 points, which is
the error bar readers should apply to any single-generation number in this
paper.
\subsection{Generalizing the \barenext{} result: a second task}
\label{sec:locksvc}

The \barenext{}-contract demonstration carried two caveats: a three-state
observable space, and 20 generations collapsing to $\sim$9 unique
solutions. To resolve both, we built a second full \barenext{} task around
\texttt{locksvc}, SysMoBench's distributed lock service, generated by PGo (a compiler from Modular PlusCal, MPCal, to Go),
chosen over the ring-buffer candidate because it runs as an ordinary Go
test with no kernel build, demonstrating in passing that the trace-capture
methodology of \S\ref{sec:traces} is not kernel-specific. The harness runs
PGo's trace-recorder test five times and folds the raw MPCal events into
observable windows over $\{\mathit{holder}, \mathit{waiters}\}$, a lock
holder plus the \emph{set} of waiting clients, yielding a 60-window
corpus (15 per action), self-consistent across runs. The set representation
is itself a correction found by the base-rate audit: a first version modeled
\texttt{waiters} as a FIFO in client-\emph{send} order and required grants
to go to the head, but the real server grants in \emph{arrival} order, which
the network reorders relative to send order (one capture logged sends
$1,3,2$ while the server queued $\langle 3,2,1\rangle$, the exact
reordering the upstream specification's NoPriorityInversion comment warns
about); the send-ordered fold silently converted ten real grants and
releases into no-op windows. Arrival order is not a function of the
client-side single steps, so at this projection the waiting clients form a
set and a grant may go to any waiter of a free lock; the corpus was
re-folded (15 no-change windows remain, all legitimately the
\texttt{CriticalSection} no-ops, identity base rate 25\%) and the study
regenerated from scratch. The fold also repairs a real trace artifact: PGo's
empty vector clocks allow a \texttt{CriticalSection} event to be logged
before its own \texttt{Grant}; the harness reorders each grant before its
critical section, which is causally sound (the critical section cannot be
entered without the grant). The replay and drivers were generalized to be
task-agnostic: per-task action domains and invariants, and the projection
rule in place of \texttt{spin}-specific keys.

The result replicates in full on the corrected corpus: \textbf{20 of 20
freshly generated \barenext{} specifications pass every phase} (5/5 per model,
including the bounded progress checks), over a 20-state observable space
with a genuine waiting set, and with 14/20 unique solution texts (4/4/4/2
per model), that is, materially different correct implementations, all
conforming, 45/45 on the state-changing windows. Both caveats are thereby
resolved: the clean sweep is not an artifact of a trivial state space, and
the models do not converge on a single memorized solution. Across the two
tasks, 40/40 model-generated \barenext{} specifications pass every phase. We
are explicit about what this number is and is not: it is evidence that the
\barenext{} contract is viable at this scale (a spinlock's 3 observable states
and a lock service's 20) and a \emph{hypothesis-generating} result about
contract minimality in general, not a settled conclusion. Both tasks
saturate, which demonstrates robustness of the contract on tasks this size
while saying nothing about tasks where a bare transition relation stops being
sufficient; the decisive test is a consensus-class protocol with non-trivial
state (Raft with logs, terms, and votes). That test is Experiment 5, and
\S\ref{sec:raft} reports where the streak ends.

\section{Experiment 5: derivation at depth on a consensus-class task}
\label{sec:raft}

The \barenext{}-contract result carried one conditionality that the first two
tasks could not discharge: both have observable relations a competent
reader can largely anticipate from the action names. A consensus protocol
does not. This experiment carries the \barenext{} contract, unchanged in shape,
to SysMoBench's \texttt{etcd} task, the Raft implementation of
etcd-io/raft, and the very system whose ``textbook'' mismodeling motivated
the benchmark~\cite{sigops2026}.

\subsection{Task and corpus}

Upstream etcd-io/raft ships its own trace instrumentation behind a build
tag, a \texttt{TraceLogger} interface whose events carry each node's
term, vote, commit index, log size, and role, added by the maintainers for
their own \tlaplus{} validation work. Our harness therefore only supplies
a sink (one JSON line per event) and determinism: five scenarios drive a
three-node cluster through the library's \texttt{RawNode} interface with
hand-delivered messages, no goroutines, and no randomness. The scenarios are
an election; replicated proposals; heartbeat rounds; a stale-log campaign
rejected by a quorum (the candidate steps down, and the incumbent
re-establishes itself at a higher term through the append reject/retry
protocol); and a higher-term re-election observed by all nodes. Delivery
orders vary across scenarios. A fold projects the event stream onto a
whole-cluster observable state,

\begin{center}
\texttt{\{nodes: \{id: \{role, term, vote, commit, log\}\}\}},
\end{center}

\begingroup\sloppy
\noindent yielding \textbf{97 windows} over the five canonical actions
(\texttt{ElectionTimeout}~8, \texttt{HandleVoteRequest}~16,
\texttt{ClientProposal}~8, \texttt{HandleAppendEntries}~57,
\texttt{HandleHeartbeat}~8). The corpus is validated the way the earlier
ones were: zero self-consistency violations, byte-identical regeneration,
and a hand-written reference \barenext{} specification that replays
\textbf{97/97} and explores cleanly, the positive control that
distinguishes ``the models failed'' from ``the fold is wrong.'' Two
projection approximations are documented with the harness (per-entry log
terms are not observable, so vote up-to-dateness reduces to an index
comparison; both are exact on these corpora by construction), and both matter below.\par\endgroup

\subsection{One-shot results}

Each of the four models generated five \barenext{} specifications from the
contract-pinned derivation prompt (state shape, action names, data
schemas; no semantics block; the transition semantics must come from
\texttt{raft.go}), scored across all phases: Phase 2/4 by the generic
bounded explorer (depth 8; roughly $1.3\times10^{5}$ distinct states, four
orders of magnitude beyond the spinlock's three), Phase 3 by the same mechanical replay as every \barenext{} study above (Table~\ref{tab:raft}).

\begin{table}[ht]
\centering
\caption{One-shot \barenext{}-contract results on \texttt{etcd} ($N{=}5$ per
model, 97-window corpus), models as columns. Every specification loads and
explores cleanly; none passes every phase.}
\label{tab:raft}
\small
\begin{tabular}{lcccc}
\toprule
 & Haiku 4.5 & Sonnet 4.6 & Opus 4.8 & Fable 5 \\
\midrule
loadable      & 5/5 & 5/5 & 5/5 & 5/5 \\
P2 clean      & 5/5 & 5/5 & 5/5 & 5/5 \\
P3 range      & 49--93 & 83--95 & 83--93 & 95--97 \\
P3 $=$ 97/97  & 0/5 & 0/5 & 0/5 & \textbf{4/5} \\
P4 hold       & 5/5 & 4/5 & 5/5 & \textbf{1/5} \\
\midrule
All-pass (all phases) & \multicolumn{4}{c}{\textbf{0/20}} \\
\bottomrule
\end{tabular}
\end{table}

The 40/40 streak ends here, and it ends informatively. What the contract
bought on the small tasks it still buys: every specification is a loadable,
explorable, replayable artifact, and nothing fails for the mechanical
reasons the SAM contract induced on the spinlock. What breaks is what no
contract shape carries.

\paragraph{The dominant failure is the vote rule, for every model.}
\texttt{HandleVoteRequest} accounts for essentially all conformance
shortfall in three models (and two windows of the fourth); Haiku
additionally mishandles append/log-match cases. The most-failed windows
(12--14 of the 20 specifications each) are the votes where the subtleties
interact: a re-election at term 3 following a rejected stale campaign, and
the second election at term 2. A representative failure: a follower at
$\{\mathit{term}{=}2, \mathit{vote}{=}\textit{self}\}$ receives a vote
request at term 3 from a candidate with an up-to-date log; the trace post
is $\{\mathit{term}{=}3, \mathit{vote}{=}\textit{candidate}\}$ (step down,
reset, grant); a failing specification produces
$\{\mathit{term}{=}3, \mathit{vote}{=}\textrm{`0'}\}$, that is, it bumps the
term and wrongly denies the vote. The projection is implicated in a
precise way: with per-entry log terms unobservable, the textbook two-field
up-to-dateness comparison is inexpressible, and a correct specification
must recognize that the faithful reduction at this projection is the index
comparison. This is the benchmark's ``textbook modeling'' failure
reproduced at protocol scale: the models write the Raft paper's rule where
the implementation-faithful reduction is required.

\paragraph{The strongest model's failure is a specification-vs-program
distinction.} Four of Fable's five specifications reach 97/97 conformance
and then fail the \texttt{CommitWithinLog} invariant under exploration.
The defect is instructive enough to quote: its heartbeat handler updates
the commit index unclamped, with a comment stating (correctly) that
a Raft leader never advertises a commit index past the follower's log.
That reasoning is sound for the closed protocol, in which only a
conforming leader generates messages. The bounded explorer, however, draws
action payloads from the declared input domains (an open-world
semantics under which a heartbeat may carry any commit value), reaches a
state with $\mathit{commit} > \mathit{log}$, and reports the violation.
(The reference specification clamps, which is also what the
implementation's own \texttt{commitTo} enforces at runtime.) Both readings
of this failure deserve stating: under the benchmark's deployed scoring
the specification fails Phase 4, full stop; read closed-world, the model
made a defensible protocol argument that happens to be the wrong argument
for a \emph{specification}, whose obligations extend to inputs the
protocol forbids. The contract never states which world \texttt{next()}
must be total over. That is the same category of paradigm commitment as
the totality-vs-enabling-conditions choice in the Phase-3 oracle
(\S\ref{sec:crosslang}), now surfacing in Phase 4, and it should be made
explicit in the contract rather than left for the strongest model to
reason past.

\paragraph{The phases re-differentiate at scale.} On the spinlock, Phases
1, 2, and 4 had no discriminating power and we reported that as a finding
about task scale. Raft confirms the other half of the prediction: bounded
exploration now catches a real defect (the unclamped commit) that the
trace corpus provably cannot (no captured trace contains a
protocol-forbidden message), while trace replay catches the vote-rule
errors that exploration alone would rate internally consistent. At
consensus scale the two external checks are complementary rather than
redundant, which is the regime the benchmark was designed for.

\subsection{One round of repair}
\label{sec:raftrepair}

Experiment 2's question transfers directly: shown exactly what failed, can
the models fix it? Each failing specification received one repair round,
consisting of its own module, every failing window with the pinned pre-state,
the real post-state, and the module's actual output, and any invariant violation with the predicate source and a violating reachable state, nothing else, and the rewrite was re-scored across all phases (Table~\ref{tab:raftrepair}).

\begin{table}[ht]
\centering
\caption{One-round repair on \texttt{etcd} (feedback: failing windows with
expected/actual, plus invariant counterexamples; one rewrite; full
re-score).}
\label{tab:raftrepair}
\small
\begin{tabular}{lccp{6.1cm}}
\toprule
Model & Baseline all-pass & Repaired all-pass & Pattern \\
\midrule
Claude Haiku 4.5  & 0/5 & 0/5 & plateau: all five at exactly 93/97, same four windows \\
Claude Sonnet 4.6 & 0/5 & 0/5 & \textbf{regression}: 3/5 collapse (95$\to$59, 93$\to$47, 95$\to$47) \\
Claude Opus 4.8   & 0/5 & 2/5 & slow convergence: two full passes, rest at 95/97 \\
Claude Fable 5    & 0/5 & \textbf{4/5} & convergence: 5/5 at 97/97; clamp fixed in 4/5 \\
\bottomrule
\end{tabular}
\end{table}

One round takes the suite from 0/20 to 6/20, all in the top two tiers, and
the spinlock's two-tier repair picture (capable models repair fully, the
weakest not at all) stratifies into four regimes.

Fable converges: every specification reaches 97/97, and in four of five
cells the invariant counterexample was sufficient for it to add the clamp
it had reasoned away, without disturbing conformance. Opus converges more
slowly: two full passes, the remainder within two windows. Haiku's
plateau is structural rather than stochastic: five independently generated
specifications all repair to exactly 93/97, failing the identical four
hard vote windows; shown its own wrong outputs on precisely those windows,
it cannot derive the rule, consistent with its non-convergence across
three rounds on the spinlock.

Sonnet's behavior is the finding we did not anticipate. In three of five
cells the rewrite fixes the quoted windows and destroys a large fraction
of previously correct behavior (95/97 to 59/97 in the worst case),
including the one invariant-failing cell, which fixed the invariant and
collapsed conformance in the same rewrite. Nothing comparable was visible
on the spinlock, where a specification is small enough that a rewrite is
effectively a regeneration. At protocol scale, \emph{preserving what
already works} emerges as its own capability axis, distinct from both
one-shot accuracy and defect repair, and a repair loop that re-scores
only the defects it fed back would report Sonnet's rewrites as successes.
Full per-window re-scoring after repair is not an optional nicety; it is
what makes the regression visible at all.

\subsection{What the boundary is}

Stated carefully: \emph{contract minimality buys transcription fidelity,
not semantic derivation.} The \barenext{} contract's 100\% held exactly as long
as the observable relation could be derived without deep reading of the
implementation; at Raft the bottleneck moves to understanding (the vote
rule under the projection, the clamp a specification owes to inputs the
protocol forbids) and no contract shape reaches it. Three
qualifications keep this within its evidence. The Raft study tests the
\barenext{} contract only; whether the SAM and \tlaplus{} arms would fare better
or worse here is untested (we would predict neither rescues the vote
rule, since the failure is language-independent modeling, but that is a
prediction, not a result). The corpus, like every corpus in this paper,
is a deterministic, protocol-easy slice: no crashes, no partitions
beyond one staged stale campaign, no reordering beyond the scheduled
variations. And the repair experiment is one round at $N{=}5$; the
regression finding in particular deserves replication before it hardens
into a claim about Sonnet rather than about this run.
\section{Scope of the cross-language comparison}
\label{sec:scope}

Before discussing the results, we consolidate in one place, rather than
scattered through the experiments, what the \jssam{}-vs-\tlaplus{}
comparison does and does not establish. Readers should carry this section's
qualifications into every cross-language claim in this paper.

\subsection{Four asymmetries}
\label{sec:asymmetries}

\begin{enumerate}
\item \textbf{Controlled and ecological are different instruments.} All
  cross-language conformance numbers come from the direct, mechanical TLC
  replay built for Experiment~4, not from the benchmark's deployed,
  agent-mediated \tlaplus{} Phase 3. The controlled path is what makes the
  languages comparable at all (an agent in the loop would confound spec
  quality with agent capability). The deployed path
  (\S\ref{sec:agentpath}) confirms the per-cause attribution (the
  harness fault vanishes, the \texttt{CHOOSE} idiom persists) but it
  scores a different window set (contention no-ops excluded rather than
  required) under an interpretive oracle (the specification's own reading
  of its action names). Its numbers and the controlled numbers answer
  different questions and must not be compared cell-to-cell.
\item \textbf{The oracles are not the same oracle.} The conformance oracle
  requires every action to be total (a successor for every pinned
  pre-state), the \barenext{}/JS paradigm by construction. Idiomatic \tlaplus{}
  writes enabling conditions and stutters; this oracle scores a disabled
  action as \emph{unscoreable}, penalizing the idiom independent of
  correctness. Every unconditional cross-language number therefore embeds a
  paradigm choice; the conditional (paradigm-fair) scores, reported
  alongside throughout, are the like-for-like reading, and under them
  the derivation-mode split disappears entirely.
\item \textbf{The \texttt{CHOOSE} trap is a guardrail matter, not a
  capability gap.} The unbounded-\texttt{CHOOSE} failures of Experiment~3
  are a known TLC anti-pattern that no experienced \tlaplus{} author
  writes, that agentic workflows already guard against~\cite{specula}, and
  that a one-line prompt guardrail eliminated in our own runs. The
  ``evaluability gap'' is a default-prompt robustness observation.
\item \textbf{One failure was ours, not the language's.} Haiku's
  \tlaplus{} Phase-2 failure was a harness fault (config generation
  produced no usable \texttt{.cfg}). It is reported per-cause everywhere,
  and the 4/4-vs-1/4 figure must not be read as purely a language effect.
\end{enumerate}

Net: what the comparison establishes is that \emph{under a matched minimal
contract and an identical mechanical oracle, JavaScript and \tlaplus{}
express this system's transition relation equally well}, and that the
observed gaps are attributable to contract shape, prompt content, an
avoidable idiom, and one harness fault. What it does not establish is any
claim about \tlaplus{} in the hands of a guarded agentic workflow, or
about the cross-language question on tasks richer than the two the
factorial studied (the Raft study of \S\ref{sec:raft} has no \tlaplus{}
arm). The \tlaplus{} figures in this paper are lower bounds.

\subsection{What \jssam{} cannot do}
\label{sec:cannotdo}

The comparison in this paper is to TLC's bounded explicit-state checking,
because that is the phase SysMoBench automates. \tlaplus{} the
\emph{ecosystem} is a strictly larger proposition, and none of the
following has any analogue in the \jssam{} backend, with either the SAM
checker or the \barenext{} explorer:

\begin{itemize}
\item \textbf{Liveness and temporal logic.} Neither explorer checks any
  LTL/CTL property. The \barenext{} explorer's bounded progress checks are
  EF-reachability within a depth bound, a bounded-horizon safety-shaped
  property, not liveness: they cannot express ``every acquisition
  eventually succeeds.''
\item \textbf{Fairness.} There is no notion of weak or strong fairness over
  actions; behaviors in which an enabled action is ignored forever are
  outside the framework's vocabulary.
\item \textbf{Unbounded verification.} Exploration is bounded by depth over
  finite intent domains. There are no refinement mappings, no inductive
  invariance proofs, no symbolic checking of parameterized or unbounded
  state spaces (cf.\ Apalache~\cite{konnov2019apalache}), and no mechanical
  theorem proving (cf.\ TLAPS~\cite{cousineau2012tlaps}).
\item \textbf{Compositional reasoning.} \tlaplus{} supports reasoning about
  a system as a composition of specifications; a \jssam{} module is a
  single closed transition system.
\end{itemize}

A \jssam{} specification is thus a \emph{checkable} artifact (bounded
safety exploration plus conformance replay against finite traces), and
the phrase ``specification language'' in this paper's title should be read
under that qualifier. On the tasks studied, the checkable subset happened
to be the only part of the pipeline with discriminating power (Phase 3),
and safety-only Phase 4 provably could not catch the headline bug
(\S\ref{sec:limitations}); that is a fact about these tasks and this
benchmark's phases, not a demonstration that the missing capabilities are
dispensable. For systems whose correctness burden lies in liveness under
fairness, which includes every consensus protocol, \tlaplus{}'s
verification capabilities have no substitute here.

\subsection{What SAM offers that the \barenext{} contract does not}
\label{sec:samoffers}

The contract-tax finding is easily over-read as ``SAM was a mistake.'' The
factorial supports a narrower statement: SAM's machinery is transcription
surface \emph{for conformance replay}, where it buys nothing, and that
is a claim about where scaffolding should be paid, not about its value.
What the SAM contract provides that a bare \texttt{next()} does not:
explicit intent signatures with finite input domains (the analogue of a
\tlaplus{} constants block, and exactly what makes Phases 2 and 4
executable without a hand-written harness); a single locus of mutation with
acceptor guards, which is the natural shape for modeling interactive and
reactive systems where proposals can be rejected; and an explorable
structure the bounded checker consumes directly. The design implication
drawn in \S\ref{sec:crosslang} is accordingly a \emph{relocation}, not a
removal: the model writes the pure transition core; the harness generates
the SAM scaffolding the checker needs, which is where TLC's apparatus
has always lived on the \tlaplus{} side.

\section{Discussion}
\label{sec:discussion}

\paragraph{``Looks right'' is not ``is right,'' in any language.}
The experiments repeatedly produced specifications that passed every
internal-consistency check while being wrong about the real system, most
vividly Fable's never-releasing spinlock, which preserved mutual exclusion
\emph{because} it was broken, and its Raft heartbeat handler, which omitted
a safety clamp on a correct protocol argument. External ground truth is
what catches both (trace replay in the first case, open-world
exploration in the second), and which check catches what turns out to
depend on task scale (\S\ref{sec:raft}). This replicates
SysMoBench's core claim~\cite{sysmobench,sigops2026} in a new language
substrate, which strengthens it: the phenomenon is about LLM modeling, not
about \tlaplus{}.

\paragraph{The familiar path is modeled; the unfamiliar path is not.}
For the three stronger models, every Experiment-1 failure lands on a
less-common code path: the non-blocking \texttt{try\_lock}-from-free
acquisition (all three) and release bookkeeping (two of the three).
Blocking acquisition, the version of a spinlock every textbook
presents, was modeled correctly by all but the smallest model. This is
the ``textbook template'' failure mode of \cite{sigops2026} reproduced at
the granularity of individual actions within a 100-line system.

\paragraph{Repair belongs in the evaluation loop, scored on the whole
specification, not the fed-back defects.}
The repair experiments suggest that single-shot leaderboard scores
understate some models (Fable, on both tasks) and overstate the ceiling of
others (Haiku, on both). Since practical formal-modeling deployments will
be agentic loops with exactly this kind of counterexample feedback,
benchmarks arguably should report both axes. The Raft repair round
(\S\ref{sec:raftrepair}) adds a methodological requirement: Sonnet's
rewrites fix the quoted defects while breaking previously correct behavior,
so a repair loop that re-checks only its own feedback would score
regressions as successes. Repair must be re-scored on the full corpus and
all phases. The repair drivers we contribute make both measurements cheap
for any \barenext{} or \jssam{} task.

\paragraph{The hypothesis, after Experiment 4: a split verdict, sharply
located.}
The practical form of the hypothesis survives: host-language familiarity plus
a self-contained executable artifact gets models to an evaluable
specification far more often (4/4 vs.\ 1/4), and evaluability is the
precondition for everything else. The strong form, that familiarity yields
more \emph{faithful} models, resolves to a tie once like is compared with
like: JavaScript in the shape of the \tlaplus{} relation matches \tlaplus{}
at 100\% for every model. (The one cross-language split, in the derivation
arms, favors \barenext{} JS and traces to the \barenext{} contract's totality
against \tlaplus{}'s guard idiom, not to the language; scored conditionally,
the tie holds there too; \S\ref{sec:limitations}.) What the three-arm data
made look like a language effect was the specification contract: every layer
of machinery between the semantics and the artifact (proposals, acceptors,
auxiliary state) is surface on which transcription slips, and the tax falls
hardest on the weakest models. Neither training-data abundance nor formality
is the operative variable on this task; \emph{contract minimality} is. The
largest single-model effect in the study is a prompt effect (Sonnet
50\%$\to$100\%), but the completed factorial shows the prompt moves only two
of four models within the SAM contract while the contract switch moves all
four to ceiling with no semantics at all: contract, then prompt, then
language is the order of what mattered here. Experiment 5 then bounds the
finding from above: at consensus scale the contract still delivers
evaluable artifacts and flawless transcription, and the residual failures
are modeling (misderived vote rules, a clamp reasoned away) which no
contract shape, in any language, was ever going to supply. The hypothesis's
own either-way framing (\S\ref{sec:intro}) lands on its second branch:
the hard part is modeling, not notation.

\paragraph{A benchmark everyone passes measures nothing, and the next
task up de-saturates it.}
There is a symmetry in the \barenext{}-contract results that deserves stating
plainly. On the two small tasks, the \barenext{} contract saturates the
benchmark: every model, Haiku included, scores 100\%, and the informative
reading is that under a minimal contract, specifying a system of that size is
a solved problem for current frontier models. What saturation exposes is
what should be measured next (derivation at depth, scale, and temporal
properties), and this study runs the first
two: at Raft scale the same contract yields 0/20, Phase 3 and Phase 4
discriminate on complementary defects, and the models separate into four
repair regimes. The benchmark's frontier moved exactly where the
saturation analysis said it would. The temporal-property axis remains open
(\S\ref{sec:cannotdo}).

\section{Limitations and threats to validity}
\label{sec:limitations}

We state these plainly; the study is a case study and should be read as one.

\begin{itemize}
\item \textbf{The controlled cross-language comparison concerns one
  system.} Experiments 1--4 all concern the spinlock, the \emph{easiest}
  class of SysMoBench task; the \barenext{}-contract validation now spans three
  tasks (\S\ref{sec:locksvc}, \S\ref{sec:raft}), but the factorial
  controlled comparison itself (the SAM and \tlaplus{} arms) has not
  been repeated beyond it, so the contract-vs-prompt ordering remains a
  small-task result. The Raft study confirmed one specific prediction (the
  phases re-differentiate at scale~\cite{sigops2026}) and bounded another
  (the \barenext{} contract's ceiling); it does not license extrapolating the
  factorial's ordering to consensus scale.
\item \textbf{Experiments 1--3 are single-generation.} Each cell in
  Tables~\ref{tab:models}--\ref{tab:tladetail} is one sample. Experiment 4's
  $N{=}5$ deployed-JS arm retroactively bounds the damage (ordering
  preserved, individual scores moving by up to $\sim$8 points) but the
  repair-loop and checkability experiments have not been replicated and
  should be read with that error bar.
\item \textbf{Trace coverage is the quiet weakness of every conformance
  number here} (audited: \texttt{scripts/trace\_coverage\_audit.py}). The
  \texttt{spin} corpus is 28 windows from \emph{two hand-authored
  deterministic scenarios}, covering 8 of the 18 observable
  (pre-state, action, data) combinations; contention is expressed only via
  \texttt{try\_lock} (2 combos), and the corpus contains \emph{zero} windows
  of a blocking \texttt{lock()} observed while the lock is held. The
  instrumentation emits the acquire event at acquisition success, where the
  pre-state is free, so the spin phase (the behavior the primitive is named
  for) is invisible in this two-variable projection by construction: a woken
  blocking waiter is indistinguishable from an uncontended acquire. The
  \texttt{locksvc} corpus is one fixed terminating workload (three one-shot
  clients; exactly 12 windows per run by protocol completion, hence the
  exactly-15-per-action count) sampled under five schedules, genuinely
  distinct (5/5 different action sequences, four different grant orders,
  including the network reordering that exposed the fold bug), but with no
  re-requests, no deeper contention, and no failure paths. The
  vector-clock reorder's causal premise is \emph{verified per event} at
  corpus build, not assumed: every \texttt{CriticalSection} step reads
  \texttt{GrantMsg} from its mailbox (message-passing happens-before), and
  the builder refuses traces where that check fails. Net: every 100\% in
  this paper means conformance to a two-variable projection of an easy,
  deterministic slice of each system's behavior, faithful to real
  executions, far from exhausting them, and the conclusion carries this
  qualifier explicitly.
\item \textbf{The \tlaplus{} comparison mixes causes.} The 1/4-vs-4/4 result
  bundles a genuine model-attributable pitfall (unbounded \texttt{CHOOSE},
  2/4) with a harness shortfall (config generation, 1/4). We report the
  attribution per model (Table~\ref{tab:tladetail}); the headline number
  should not be read as purely a language effect. The repair asymmetry here, raised in review by Erik Meijer~\cite{meijer2026}, is closed in full: one round of
  TLC-feedback repair fixes both \texttt{CHOOSE} failures, and the
  repaired specifications hold all seven invariants through the pipeline
  (\S\ref{sec:vstla}).
\item \textbf{The derivation-level comparison splits
  the languages for the first time.} Experiment 4 as first designed
  measured transcription, not derivation: it had no
  \tlaplus{}-without-semantics arm to pair against the \barenext{} one. That arm
  (\texttt{tld}: the constrained \tlaplus{} contract with the
  semantics block removed, mirroring the \barenext{} derivation arm) completes
  the design. In
  derivation mode the \barenext{} JS contract beats \tlaplus{} for Opus and Sonnet
  ($\Delta{=}.214$, $p{=}.0079$ each; 100\% vs.\ 78.6\%), directionally for
  Haiku (87.1\%, $p{=}.17$), and ties for Fable (100\%). The mechanism is
  precise and is not wrong post-states: every \tlaplus{} shortfall window is
  \emph{unscoreable}. Without instruction, most models write the classic
  \tlaplus{} idiom in which \texttt{AcquireLock} is \emph{guarded} on the
  lock being free, so a failed attempt is a disabled action rather than an
  observable no-op step, and the pinned contention windows have no successor.
  The \barenext{} contract's \emph{totality} (\texttt{next()} must return a state)
  structurally forbids this; \tlaplus{}'s idiom permits it, and the
  semantics block's ``held $\Rightarrow$ unchanged'' sentence turns out to
  be what fixes it ($\texttt{tld}$ vs.\ $\texttt{tla}$: $p{=}.0079$ for both
  affected models). Under the pre-registered \emph{conditional} scoring the
  guard-idiom specs are 100\% correct on every window they can replay, so
  the shortfall is replay-contract incompatibility, not incorrect modeling,
  and both numbers are reported, with the unconditional as headline.
  ``Contract first, prompt second'' thus refines to: \emph{the prompt
  matters exactly where the contract under-constrains}.
\item \textbf{The plain-JS arm resolves the substrate asymmetry, but was
  added post hoc and answers a narrower question.} The three-arm design left
  ``executable machinery admits bugs a two-variable relation cannot'' as an
  open alternative; the fourth arm was designed to test exactly that, after
  the three-arm results were known (its data were then collected by the same
  blind mechanical replay). Its conclusion has since been extended rather
  than bounded: a generic \texttt{\{init, next\}} explorer replaces the SAM
  scaffolding for Phases 2 and 4 (at equivalent bounded-safety checking
  power on these state spaces), and 20/20 model-generated \barenext{}
  specifications pass every phase on \texttt{spin}, but that
  demonstration inherits \texttt{spin}'s three-state observable space and a
  converged solution set ($\sim$9 unique among 20), so the finding needed a
  second task. The \texttt{locksvc} study (\S\ref{sec:locksvc}) supplies it
  (a 20-state space, 14/20 unique solution texts, 20/20 passing),
  generalizing the contract-tax removal across tasks of this scale, though
  not yet to large distributed protocols.
\item \textbf{$b{=}0$ is partly structural, and the effective sample is
  small.} When the \tlaplus{} arm passes every window it fails none, so a
  JS-only pass is impossible by construction; any paired comparison reduces
  to whether the \jssam{} failure count is significant. Moreover the
  generations collapse to 1--2 unique behavioral fingerprints per (model,
  arm): a pooled window-level McNemar would count the
  same defect up to five times, so the analysis of record is the
  generation-level permutation test of
  Table~\ref{tab:mcnemar}. At the strictest unit (the unique solution,
  $n{=}1$--$2$ per cell) no significance test is meaningful; the
  per-solution effect sizes stand on their own.
\item \textbf{The direct TLC replay is the controlled condition, not the
  deployed one.} The benchmark ships an agent-mediated \tlaplus{} Phase 3;
  pinning spec variables to the trace schema trades realism for
  comparability. Whether a ``pinned-schema'' Phase-3 mode should exist as a
  first-class controlled condition alongside the agent path is a
  methodological question we are putting to the maintainers.
\item \textbf{Phase 4 for \jssam{} is safety-only and could not fail the
  headline bug in principle.} This is stronger than a template-count
  difference (seven \tlaplus{} templates vs.\ three \jssam{}). The three
  \jssam{} invariants are all safety properties over reachable states, and a
  never-releasing lock satisfies every one of them \emph{because} it is
  broken: mutual exclusion holds trivially when the lock never changes
  hands, status consistency holds for a permanently locked holder, and the
  shipped \texttt{NoDeadlock} (``some thread is not \texttt{trying}'') is
  satisfied by the stuck holder itself being \texttt{locked}. We verified
  this empirically: a release-is-a-no-op mutant of the known-good
  specification passes Phase 4 with the identical 3/3 verdict as the correct
  one. The properties that would discriminate (every waiter eventually
  acquires, every hold is eventually released) are liveness, and neither
  the SAM checker nor the \barenext{} explorer checks any temporal property (the
  \jssam{} template library states the exclusion explicitly). Fable's defect
  is doubly invisible: it manifests only under a pinned observable
  pre-state, so even a liveness-capable from-init checker would pass its
  specification; only trace replay catches it. As mitigation the \barenext{}
  explorer now supports \emph{bounded progress} checks (EF-reachability:
  from every reachable state satisfying a premise, a goal state must be
  reachable within the bound, not liveness: no fairness, bounded
  horizon). These fail the never-releasing mutant
  (\texttt{ReleaseProgress}) and would fail the inert Haiku specification
  (\texttt{AcquireProgress}) that safety-only Phase 4 passed vacuously; all
  40 saved model-generated \barenext{} specifications pass them on both tasks.
\item \textbf{Model family.} All four models are from one vendor. The
  benchmark's configuration already includes other vendors' models; the
  comparison should be broadened before generalizing. A further
  reproducibility caution: several of the exact model versions used here
  may not be externally accessible; the saved specifications and raw
  per-window data, not the model identifiers, are the reproducible objects.
\item \textbf{Missing arms, with status.} Several
  acknowledged-but-load-bearing gaps remain; their current state: the
  \emph{derivation arms} (no semantics block, both languages) are run
  (the one gap that, once filled, changed a conclusion);
  \emph{multi-round repair} is run on \texttt{spin} (Haiku does not
  converge; \S\ref{sec:repair}) and single-round on \texttt{etcd}
  (\S\ref{sec:raftrepair}); the \emph{as-deployed agent-mediated
  \tlaplus{} Phase 3} is run (\S\ref{sec:agentpath}) and validates
  the attribution but, as an interpretive oracle over a different window
  set, does not substitute for the controlled comparison;
  \emph{other vendors} remain blocked on credentials (the registry entries
  exist in \texttt{config/models.yaml}; only Anthropic keys are available
  in this environment); and the \emph{factorials on \texttt{locksvc} and
  \texttt{etcd}} remain open (the \barenext{} arms exist; the SAM and \tlaplus{}
  arms need per-task prompts and TLC replay modules). Each is inventoried
  with run instructions in \texttt{REPLICATION.md}.
\item \textbf{The Raft study's own limits.} The corpus is five
  deterministic scenarios with no crashes, no partitions beyond one staged
  stale campaign, and no unscheduled reordering; the projection omits
  per-entry log terms, with two documented approximations that are exact
  on these corpora by construction (\S\ref{sec:raft}); richer corpora
  could separate specifications these cannot. The repair round is one
  round at $N{=}5$: the four-regime stratification, and Sonnet's
  regression in particular, should be read as this run's result pending
  replication. And the open-world reading of the \texttt{CommitWithinLog}
  failures is a scoring commitment: we report deployed (open-world)
  numbers as headline and the closed-world reading alongside, mirroring
  the conditional-scoring convention of Experiment 4. Two cells initially
  recorded as unloadable were traced to a transient sandbox failure under
  machine load and re-scored clean; the raw results file preserves the note.
\item \textbf{Author positionality, and the case for independent
  replication.} The first author is the author of the SAM pattern. The
  evaluation pipeline is automated and the corpus is model-independent,
  which limits (but does not eliminate) the room for favorable design
  choices; and the study's sharpest negative finding, the contract-tax
  attribution, cuts \emph{against} the SAM contract as deployed, which
  we take as evidence the pipeline is not tilted toward it. But the deeper
  issue is that the load-bearing materials on
  \emph{both} sides of every comparison are experimenter-authored: the
  trace scenarios, the semantics block, the derivation prompts, and the
  repair prompt. No amount of automation neutralizes that. The repository
  therefore ships a replication guide (\texttt{REPLICATION.md}) that
  reruns every analysis from committed raw data without API access,
  regenerates every experiment from pinned prompts and models, and
  inventories each experimenter-authored material as an independent
  variable to audit or vary. We consider an independent replication by
  researchers unaffiliated with SAM (ideally the SysMoBench
  maintainers) a prerequisite for treating the cross-language
  conclusions as more than a well-documented single-author case study.
\end{itemize}
\section{Related work}
\label{sec:related}

\paragraph{Benchmarking LLM formal modeling.}
SysMoBench~\cite{sysmobench,sigops2026} is the direct foundation of this
work: it contributes the four-phase methodology, the eleven-system task
suite, and the finding that conformance, not syntax, is where LLM
specifications fail. Our contribution is orthogonal: holding the benchmark
fixed and varying the specification \emph{language}, including the first
non-formal one. The Specula agent~\cite{specula} shows that agentic
workflows can saturate current SysMoBench tasks in \tlaplus{}; our repair
experiment probes the smallest unit of such a workflow (one counterexample
round) across models.

\paragraph{The wider LLM$\times$formal-methods field.}
A relevance-ranked sweep of the 2022--2026 literature (40 works; the map
ships with the artifacts) places this study in a thin slice of a
fast-moving field. The center of gravity is autoformalization and theorem
proving (Lean/Isabelle proof generation and
repair~\cite{wu2022autoformalization,yang2023leandojo,first2023baldur}),
followed by loop-invariant and precondition generation for deductive
verification~\cite{chakraborty2023ranking}. Specification-focused work
concentrates on natural-language-to-temporal-logic
translation~\cite{cosler2023nl2spec} and declarative-specification
repair~\cite{ese2025repair}; benchmarks are proliferating (e.g., OSVBench
for OS-verification specification generation~\cite{osvbench2025}) but grade
against reference specifications, proof success, or human judgment rather
than execution-trace ground truth. Against that corpus, this study's two
distinctive commitments, conformance to traces captured from the running
system as the discriminating oracle and a controlled
contract$\times$prompt$\times$language factorial, are both minority
positions. The closest contemporaneous works, an NL-to-\tlaplus{}
correctness evaluation and counterexample-driven \tlaplus{}
repair~\cite{nl2tla2026,tracefix2026}, each parallel a single axis of the
design here (one-shot generation; repair); the repair-as-separate-axis
finding of Experiment 2 matches an emerging pattern across proof
repair~\cite{first2023baldur} and specification
repair~\cite{ese2025repair}.

\paragraph{Trace conformance for specifications.}
Validating specifications against implementation traces has industrial
precedent: MongoDB's eXtreme modeling checked \tlaplus{} models against
driver traces~\cite{davis2020extreme}, and trace validation is the standard
answer to ``does the spec match the code''~\cite{sysmobench}. SysMoBench's
transition-window formulation, which we adopt unchanged, makes conformance
per-action and per-window rather than whole-trace.

\paragraph{Formal methods practice.}
\tlaplus{}~\cite{lamport2002specifying} with TLC~\cite{yu1999tlc},
Alloy~\cite{jackson2002alloy}, and CSP-based PAT~\cite{sun2009pat} are the
benchmark's formal backends; industrial deployments
(e.g.,~\cite{newcombe2015amazon}) motivate the whole enterprise. SAM
\cite{sam,dubray2016infoq} occupies an unusual position: an engineering
pattern with \tlaplus{}-derived semantics, here repurposed as a
specification language precisely because it sits at the intersection of
``formally disciplined'' and ``abundant in training data.''

\paragraph{Executable specifications are not new.}
The formal-methods community has long built specification languages that
are executable while remaining formally checkable, and this paper's
contribution must be located against that lineage rather than presented as
discovering executability. Event-B with Rodin supports executable formal
modeling under refinement~\cite{abrial2010eventb}; Dafny is an imperative
language with specifications and automated verification built
in~\cite{leino2010dafny}; Frama-C/ACSL attaches checkable contracts to
executable C~\cite{kirchner2015framac}; refinement/liquid types make
specifications part of an executable program's type
structure~\cite{rondon2008liquid}; and within the \tlaplus{} family itself,
PlusCal~\cite{lamport2009pluscal} and Quint~\cite{quint} deliberately trade
notation toward the programmer. The \tlaplus{} ecosystem likewise extends
well past TLC, the only tool compared here: TLAPS provides mechanical
proof~\cite{cousineau2012tlaps} and Apalache symbolic model
checking~\cite{konnov2019apalache} (\S\ref{sec:cannotdo}). What is new in
this study is not executable specification but its \emph{LLM-generation}
angle under a trace-grounded oracle: whether frontier models produce more
faithful specifications in an executable mainstream language than in a
formal one, measured by conformance to traces of the running system,
and the finding that the specification contract, not the language,
dominated that outcome.

\section{Conclusion and future work}
\label{sec:conclusion}

We extended SysMoBench with \jssam{}, its first non-formal specification
backend, built the kernel trace-capture pipeline needed to give it a real
conformance phase, and ran five experiments across three tasks: the
Asterinas spinlock, a distributed lock service, and Etcd Raft.
The results make five points with unusual precision for a case study.
On the small tasks, transition validation against real traces is the only
phase that discriminated among four frontier models (0--86.4\% on
state-changing windows, with all other phases unanimous and the weakest
model sitting exactly at the corpus's identity base rate). One round of
trace-feedback repair reveals a self-correction axis that one-shot scores
conceal: capable models repair fully and verifiably (a held-out audit found
root-cause fixes, not memorized windows), the weakest not at all, and at
protocol scale that axis stratifies further into convergence, slow
convergence, regression, and plateau. The same models that reach an
evaluable specification in JavaScript 4/4 times manage it in \tlaplus{} 1/4
times, for two identified causes, one of them a harness fault, both later
removed entirely by a one-line prompt guardrail, both confirmed per-cause by
a run of the benchmark's deployed agent-mediated path, and both
\texttt{CHOOSE} failures repaired by one round of TLC-feedback repair: a
default-prompt robustness gap, not a language-capability one. Under a
pre-registered, paired comparison completed to a full $3{\times}2$
factorial in (contract, prompt), the faithfulness question resolves in an
unexpected place: the conformance gap is attributable to the SAM contract's
machinery, not to JavaScript, not to formality, and not primarily to prompt
prescriptiveness. Plain JavaScript in the shape of the \tlaplus{} relation
ties \tlaplus{} at 100\% for every model, with or without spec-level
semantics in the prompt. And the winning contract, carried to a consensus
protocol, finds its boundary: 0 of 20 one-shot specifications pass every
phase, for reasons that are modeling rather than transcription, with one
repair round recovering 6 of the 20.

The hypothesis that motivated the backend is neither confirmed nor dead; it
is answered more usefully than either. Stated plainly, on the tasks tested
the completed study ranks what governed LLM specification quality:
\emph{contract shape first, prompt second, language nowhere}, a ranking
established on the spinlock factorial and not yet replicated on a richer
task. Familiarity buys \emph{checkability};
contract minimality buys \emph{transcription fidelity}, and, Experiment
5 shows, only that: on the spinlock every model derives the observable
semantics unaided when the target is a bare transition function, but at
consensus scale the bottleneck moves to semantic derivation itself (the
vote rule under the projection, the clamp a specification owes to inputs
the protocol forbids) and no contract shape reaches it. The prompt is the
model-dependent factor: four lines of semantics moved Sonnet 50\%$\to$100\%
and Fable 57.9\%$\to$95.7\% within the SAM contract, but moved Opus and
Haiku insignificantly, and at the \barenext{} contract it had nothing left to
move (a ceiling at which effect ordering among the perfect cells is, we
note, not identifiable). Every layer of ceremony between the semantics and
the artifact cost conformance; the language itself, once shape was matched,
contributed nothing measurable (the derivation-mode split between \barenext{}
JS and \tlaplus{} traces to contract totality, not language: scored
conditionally, the guard-idiom \tlaplus{} specifications are perfect on
every scoreable window). On the practical question, whether JavaScript can
replace \tlaplus{}, the evidence splits cleanly. For LLM-generated
conformance modeling at this scale: yes, interchangeable on fidelity
\emph{on what we measured}, which is conformance to a two-variable
observable projection of deterministic, easy slices of each system's
behavior (\S\ref{sec:limitations}); neither language has been tested on
uncovered combinations, projection-invisible behavior such as the blocking
spin, richer workloads, or adversarial interleavings. Within that measured
slice the tie is exact, and JavaScript carries fewer operational failure
modes (no \texttt{CHOOSE} trap, no configuration artifact). For
verification: no. Liveness, fairness, temporal-logic properties,
large-state and unbounded checking, and mechanical proof (TLC's temporal
checking, Apalache, TLAPS) have no equivalent in either bounded safety
explorer used here (the SAM library's checker or its 40-line generic
replacement), and remain \tlaplus{}'s distinctive value
(\S\ref{sec:cannotdo}). The demonstrated role for executable JavaScript is
as a checkable artifact for bounded model checking and trace-conformance
validation, complementing, not replacing, \tlaplus{}'s verification
capabilities. The search this motivates is therefore not for a better
specification \emph{language} but for a better specification
\emph{contract}: \texttt{\{init, next\}} over observable keys, in whatever
language the model writes well, with richer machinery generated by the
harness rather than transcribed by the model. Two questions that the
small-task results left open, derivation at depth and scale, are answered
here for the \barenext{} contract, and the answer reshapes the agenda:
derivation at depth is where one-shot generation fails and where feedback
loops earn their keep. What remains genuinely open is the
\emph{cross-language} question at that depth (whether \tlaplus{} one-shot
generation fares any better on Raft, which we predict it does not, since the
failures are modeling, but it is a prediction), the temporal properties no
bounded safety explorer checks, and corpora adversarial enough to separate
specifications these deterministic slices cannot.

The experiments that follow are concrete: (1)~promoting the \barenext{}
specification shape from demonstration to backend, and, following
Experiment 5, making the open-world totality obligation explicit in the
contract: a specification's \texttt{next()} answers for the declared
domain, not only for protocol-reachable inputs; (2)~the cross-language
factorials on \texttt{locksvc} and \texttt{etcd} (the \barenext{} arms exist; the
SAM and \tlaplus{} arms are scoped in the replication guide), the test
of whether ``language nowhere'' survives depth; (3)~a
prompt-prescriptiveness
dose--response study, treating the semantics block as the manipulated
variable rather than a nuisance; (4)~multi-round repair on \texttt{etcd},
to establish whether Opus converges, whether Sonnet's regression is
systematic, and where Fable's fifth cell lands; (5)~hardening the
benchmark surfaces this study kept tripping on, the configuration
generator and the invariant pipeline's name-collision behavior
(\S\ref{sec:vstla}), both now attributed per-cause; and (6)~replication
beyond one model family. We are contributing the backend, harnesses, drivers, corpora, and
the controlled-comparison methodology upstream so these can be run, and
challenged, by the community. One such test has since been reported~\cite{finixpos}: a study of production software, the payment workflow of an operational point-of-sale system that keeps the register, payment terminal, and payment processor in agreement, runs the same \jssam{}-vs-\tlaplus{} comparison and finds the contract-structure ordering replicating across seven models from two vendors.

\section*{Acknowledgments}
We thank the SysMoBench / Specula maintainers for the benchmark, the
integration guidance that shaped the \jssam{} backend's scope, and the
reference instrumentation for the Asterinas spinlock.

\section*{Reproducibility}
\begingroup\sloppy
All artifacts are public: the \jssam{} backend
(\texttt{tla\_\allowbreak eval/\allowbreak languages/\allowbreak js\_\allowbreak sam.py}, \texttt{tools/\allowbreak js-sam/\allowbreak }), the trace
harness (\texttt{scripts/\allowbreak harness/\allowbreak spin/\allowbreak }), the captured corpus
(\texttt{data/\allowbreak sys\_\allowbreak traces/\allowbreak spin/\allowbreak }), the repair driver
(\texttt{scripts/\allowbreak repair\_\allowbreak phase3.py}), the Experiment-4 apparatus (the
constrained and plain-JS prompts, the direct TLC replay and its
functional-check audit (\texttt{scripts/\allowbreak tla\_\allowbreak direct\_\allowbreak tv.py}, the 20 audited
specifications in \texttt{output/\allowbreak tla\_\allowbreak specs/\allowbreak },
\texttt{output/\allowbreak tla\_\allowbreak functional\_\allowbreak audit.json}, the permissive control in
\texttt{tools/\allowbreak tla/\allowbreak }), the paired factorial driver with raw
per-window data (\texttt{scripts/\allowbreak tla\_\allowbreak phase3\_\allowbreak study.py},
\texttt{output/\allowbreak tla\_\allowbreak phase3\_\allowbreak study.json}), the generation-level analysis of
record and the repair-generalization audit
(\texttt{scripts/\allowbreak tla\_\allowbreak phase3\_\allowbreak analysis.py},
\texttt{scripts/\allowbreak repair\_\allowbreak generalization.py}), the trace-coverage audit
(\texttt{scripts/\allowbreak trace\_\allowbreak coverage\_\allowbreak audit.py}), the pre-registered design and
results (\texttt{docs/\allowbreak js\_\allowbreak sam\_\allowbreak tla\_\allowbreak phase3\_\allowbreak *.md})), the \barenext{}-contract
prototype and full-pipeline studies (\texttt{tools/\allowbreak plain-js/\allowbreak },
\texttt{scripts/\allowbreak plain\_\allowbreak js\_\allowbreak tv.py},
\texttt{scripts/\allowbreak lean\_\allowbreak full\_\allowbreak pipeline\_\allowbreak study.py}, the 40 saved generations in
\texttt{output/\allowbreak lean\_\allowbreak specs*/\allowbreak }, \texttt{docs/\allowbreak js\_\allowbreak sam\_\allowbreak lean\_\allowbreak contract.md}),
the \texttt{locksvc} harness and corpus
(\texttt{scripts/\allowbreak harness/\allowbreak locksvc/\allowbreak }, \texttt{data/\allowbreak sys\_\allowbreak traces/\allowbreak locksvc/\allowbreak }),
the \texttt{etcd} harness, corpus, reference specification, and study
artifacts (\texttt{scripts/\allowbreak harness/\allowbreak etcd/\allowbreak }, \texttt{data/\allowbreak sys\_\allowbreak traces/\allowbreak etcd/\allowbreak }
with pinned provenance, \texttt{output/\allowbreak lean\_\allowbreak specs\_\allowbreak etcd\*/\allowbreak },
\texttt{output/\allowbreak lean\_\allowbreak full\_\allowbreak pipeline\_\allowbreak etcd.json},
\texttt{output/\allowbreak lean\_\allowbreak etcd\_\allowbreak repair.json},
\texttt{scripts/\allowbreak repair\_\allowbreak lean\_\allowbreak etcd.py}), the deployed agent-path runs
(\texttt{scripts/\allowbreak launch\_\allowbreak tv\_\allowbreak eval.sh}; per-spec reports in
\texttt{output/\allowbreak agent\_\allowbreak tv\_\allowbreak reports/\allowbreak }), the \tlaplus{} repair round
(\texttt{scripts/\allowbreak repair\_\allowbreak tla\_\allowbreak spin.py}, repaired specifications and TLC
output in \texttt{output/\allowbreak tla\_\allowbreak repair/\allowbreak }), the methodology note
(\texttt{paper/\allowbreak methodology.md}), the literature map
(\texttt{paper/\allowbreak literature\_\allowbreak map.md}), and the experiment
write-ups
(\texttt{docs/\allowbreak js\_\allowbreak sam\_\allowbreak *.md}). All of the above reside on a single branch, \texttt{js-sam-tla-phase3}, of the SysMoBench fork at \url{https://github.com/jdubray/SysMoBench-1}; the \jssam{} backend itself is additionally merged upstream into \url{https://github.com/specula-org/SysMoBench} (PR~\#17), with the cross-platform and spin trace-harness changes proposed in PR~\#21.
Requirements: Docker, Node $\geq$ 20, Python 3, and (for the \tlaplus{} arm)
Java with \texttt{tla2tools}. A single script reproduces the kernel capture:
\texttt{scripts/\allowbreak harness/\allowbreak spin/\allowbreak run.sh}. A replication guide
(\texttt{REPLICATION.md}, repository root) re-derives every analysis from
committed raw data without API access, regenerates every experiment from
pinned prompts and models, and inventories the experimenter-authored
materials as independent variables to audit or vary.

\endgroup

\end{document}